\documentclass[useAMS,usenatbib]{mn2e}

\usepackage{graphicx}
\usepackage{amsmath}
\usepackage{amsfonts}
\usepackage{amssymb}
\usepackage{rotating}
\usepackage[english]{babel}

\title{On  the  possible  observational   signatures  of  white  dwarf
  dynamical interactions}

\author[G. Aznar--Sigu\'an et al.]{G. Aznar--Sigu\'an$^{1,2}$,
                                E. Garc\'\i a--Berro$^{1,2}$,
				M. Magnien$^{1,3}$ and 
                                P. Lor\'en--Aguilar$^{4}$\\
       $^1$Departament de F\'\i sica Aplicada, 
           Universitat Polit\`ecnica de Catalunya,
           c/Esteve Terrades 5, 
           08860 Castelldefels, Spain\\
       $^2$Institute for Space  Studies of Catalonia,
           c/Gran Capit\`a 2--4, Edif. Nexus 104,   
           08034  Barcelona, Spain\\
       $^3$\'Ecole Polytechnique F\'eminine, 
           3 bis, rue Lakanal
           92330 Sceaux,
           France\\
       $^4$School of Physics, University of Exeter, 
           Stocker Road, Exeter, 
           UK EX4 4QL, United Kingdom}

\begin{document}

\date{\today}

\maketitle

\begin{abstract}
We  compute  the  possible  observational signatures  of  white  dwarf
dynamical interactions in dense stellar environments. Specifically, we
compute the  emission of gravitational  waves, and we compare  it with
the  sensitivity  curves  of planned  space-borne  gravitational  wave
detectors. We also compute the  light curves for those interactions in
which a detonation occurs, and one  of the stars is destroyed, as well
as the corresponding neutrino luminosities. We find that for the three
possible outcomes of these interactions --- which are the formation of
an eccentric binary system, a  lateral collision in which several mass
transfer episodes occur, and a direct  one in which just a single mass
transfer  episode takes  place ---  only those  in which  an eccentric
binary  are  formed   are  likely  to  be  detected   by  the  planned
gravitational wave mission eLISA, while more sensitive detectors would
be able to  detect the signals emitted in lateral  collisions.  On the
other hand,  the light curves  (and the thermal neutrino  emission) of
these  interactions are  considerably different,  producing both  very
powerful  outbursts  and  low  luminosity events.   Finally,  we  also
calculate  the X-ray  signature  produced in  the  aftermath of  those
interactions  for which  a merger  occurs. We  find that  the temporal
evolution follows a power law with the same exponent found in the case
of  the  mergers of  two  neutron  stars,  although the  total  energy
released is smaller.
\end{abstract}


\section{Introduction}

Close encounters of two white dwarfs in dense stellar environments, as
the central regions of galaxies or the cores of globular clusters, are
interesting phenomena  that have  several potential  applications, and
hence deserve  close scrutiny.  This  type of interactions can  be the
result  of either  a serendipitous  approach due  to the  high stellar
density  of  the  considered  stellar  system or  the  result  of  the
interaction of a  binary system containing a white dwarf  with a third
star  in  a triple  system,  via  the  Kozai-Lidov mechanism  ---  see
\cite{Kushnir2013}  for  a  detailed  description  of  this  scenario.
Nevertheless,  the details  of how  the intervening  white dwarfs  are
brought close enough to experience  such dynamical interactions do not
play  an important  role  in the  result of  the  interaction, so  the
outcome of  the interaction is  independent of these details,  and the
gross features of the hydrodynamical evolution are totally general.

Among the possible applications of  these interactions we mention that
it has been  shown \citep{Rosswog2009} that head-on  collisions of two
white  dwarfs is  a viable  mechanism  to produce  Type Ia  Supernovae
(SNIa), one  of the most energetic  events in the Universe.   This, in
turn, has  obvious implications in Cosmology,  as SNIa are one  of the
best distance indicators.  However, there  are situations in which the
two white  dwarfs interact but  the interaction  does not result  in a
powerful thermonuclear outburst leading  to the complete disruption of
the two stars, whereas in some  other the release of nuclear energy is
very modest,  and finally there  are other  cases in which  no nuclear
energy is released  at all.  Specifically, it has  been recently shown
\citep{Aznar2013}  that white  dwarf close  encounters can  have three
different  outcomes,  depending  on  the  initial  conditions  of  the
interaction -- namely, the initial  energy and angular momentum of the
pair of white dwarfs, or equivalently the impact parameter and initial
velocity of the pair of  stars.  In particular, these interactions can
lead to  the formation  of an  eccentric binary  system, to  a lateral
collision  in  which  several   mass  transfer  episodes  between  the
disrupted  less massive  star  and  the more  massive  one occur,  and
finally to  a direct  collision, in which  only one  catastrophic mass
transfer event occurs.  Additionally, it  turns out that for these two
last outcomes  there is a  sizable region  of the parameter  space for
which the  conditions for a  detonation to  occur are met,  leading in
some cases  to the  total disruption  of both white  dwarfs, and  to a
large   dispersion  of   the  nuclear   energy  released   during  the
interaction.

According to the  above mentioned considerations, during  the last few
years the  close encounters of  two white  dwarfs is a  research topic
that  has been  the subject  of renewed  interest.  Specifically,  the
dynamical  behavior  of  these  interactions  has  been  simulated  by
\cite{Rosswog2009},        \cite{Raskin2009},        \cite{Loren2010},
\cite{Raskin2010},     \cite{Hawley2012},     \cite{Aznar2013}     and
\cite{Kushnir2013}. It is to be noted that in almost all the cases the
simulations  were carried  out using  Smoothed Particle  Hydrodynamics
(SPH)  codes, given  the  intrinsic difficulties  of simulating  these
encounters  on   an  Eulerian  grid.   Nevertheless,   most  of  these
simulations  have  focused on  the  head-on  collisions of  two  white
dwarfs, and  the only  works in which  a systematic  and comprehensive
study of the effects of the  initial conditions on the result of these
interactions   was   done   are    those   of   \cite{Loren2010}   and
\cite{Aznar2013}.    In  particular,   \cite{Loren2010}  studied   the
dynamical  interactions  of two  otherwise  typical  white dwarfs,  of
masses $0.6\,  M_{\sun}$ and  $0.8\, M_{\sun}$, for  a reduced  set of
energies and angular momenta.  Later on, \cite{Aznar2013} extended the
previous  work,  studying a  significantly  broader  range of  initial
conditions for  a larger set of  white dwarf masses and  core chemical
compositions, including  helium, carbon-oxygen, and  oxygen-neon white
dwarfs.

The  study  of  \cite{Aznar2013}   demonstrated  that  the  masses  of
$^{56}$Ni synthesized in the strongest dynamical interactions can vary
by orders of magnitude, and so do the typical luminosities.  Actually,
the  mass of  $^{56}$Ni obtained  in the  several sets  of simulations
performed so far, and thus  the corresponding luminosities, range from
those typical of macro-nova events, to sub-luminous and super-luminous
supernova events.   Nonetheless, it is  important to realize  that the
spread  in the  mass of  $^{56}$Ni  produced during  the most  violent
phases of the most energetic interactions  does not only depend on the
adopted initial conditions but also,  as shown by \cite{Aznar2013} and
\cite{Kushnir2013},  on the  adopted  prescription  of the  artificial
viscosity, and  on the resolution  employed in the  simulations, among
other (less  important) factors.  Additionally,  one has to  note that
the parameter  space of  such dynamical interactions  is huge  and the
complexity of the physical mechanism  of the interactions is extremely
high.   \cite{Aznar2013}  chose to  account  for  the first  point  by
producing a vast  set of numerical simulations, but this  comes of the
expense  of the  second issue.   Consequently, most  likely the  tidal
interaction   is  reasonably   captured  in   moderately-resolved  SPH
simulations, but combustion processes are possibly not completely well
resolved. Nevertheless, although a  thorough resolution study is still
lacking,  the   masses  of   $^{56}$Ni  synthesized  in   the  several
independent sets of simulations agree within a factor of $\sim 5$, and
the  qualitative  picture of  the  hydrodynamical  evolution is  quite
similar in  all the  cases --- see  Sect.~5.1 of  \cite{Aznar2013} for
additionail details.

Despite the  large efforts  invested in modelling  these interactions,
quite  naturally most  of them  have focused  on the  dynamics of  the
events, and on the  resulting nucleosynthesis, whilst little attention
has been paid to other  possible observational signatures.  This is in
contrast with the  considerable deal of work done so  far for the case
in which  two neutron stars  merge or  collide --- see,  for instance,
\citep{Piran2013}, and  references therein.  In  this paper we  aim at
filling  this  gap  by  computing  the  gravitational  waveforms,  the
corresponding light curves for those events in which some $^{56}$Ni is
synthesized,  the  associated emission  of  neutrinos,  and the  X-ray
luminosity of  the fallback  material.  All  of them  individually, or
used  in  combination,  would  hopefully allow  us  to  obtain  useful
information about these events.
  
The emission  of gravitational  waves in  these iteractions  should be
quite  apparent,  given that  the  two  interacting white  dwarfs  are
subject to  large accelerations,  and that in  most of  the situations
there is no symmetry.  Although  gravitational waves have not been yet
detected, with  the advent  of the  current generation  of terrestrial
gravitational wave detectors and of space-borne interferometers, it is
expected  that the  first  direct  detections will  be  possible in  a
future.  In particular,  much hope has been placed  on the space-based
interferometer eLISA,  a rescoped version  of LISA, which  will survey
for the  first time  the low-frequency  gravitational wave  band (from
$\sim$0.1~mHz to $\sim$1~Hz).  The  timescales of the close encounters
of two white  dwarfs correspond precisely to  this frequency interval.
Also, for those  events in which an explosive behavior  is found it is
clear that some information about  the dynamical interactions could be
derived from  the analysis of  the light  curves.  On the  other hand,
neutrino  emission in  these events  should be  noticeable, given  the
relatively high  temperatures achieved during the  most violent phases
of  the  interaction  ($\sim  10^9$~K), and  an  assessment  of  their
detectability  is  lacking.   Finally,  the X-ray  luminosity  of  the
fallback material interacting with the disk resulting in the aftermath
of the  interactions would also  eventually help in  identifying these
events.

The  paper is  organized as  follows.   We first  discuss the  methods
employed to  characterize the observational signatures  of white dwarf
close    encounters    and    collisions    (Sect.~\ref{sec:methods}).
Sect.~\ref{sec:results}  is  devoted to  analyze  the  results of  our
theoretical          calculations.           Specifically,          in
Sect.~\ref{sec:results:GWR}  we study  the  emission of  gravitational
waves.  It  follows Sect.~\ref{sec:results:LCs}  where we  discuss the
late-time light curves for those interactions which result in powerful
detonations,  while  in   Sect.~\ref{sec:results:nu}  we  present  the
thermal  neutrino  fluxes  and  we  assess  their  detectability.   In
Sect.~\ref{sec:results:Xray},  we  compute the  fallback  luminosities
produced  in the  aftermath of  those interactions  which result  in a
central    remnant    surrounded    by   a    disk.     Finally,    in
Sect.~\ref{sec:conclusions} we summarize our most relevant results, we
discuss their significance, and we draw our conclusions.

\section{Numerical setup}
\label{sec:methods}

In  this  section   we  describe  how  we  compute   the  emission  of
gravitational waves, the  light curves for those events  which have an
explosive  outcome,  the  neutrino  fluxes,  and  the  fallback  X-ray
luminosity of  the remnants,  for those cases  in which  the dynamical
interaction  is not  strong enough  to  disrupt both  stars.  All  the
calculations presented  in Sect.~\ref{sec:results}  are the  result of
post-processing the  SPH calculations  of interacting white  dwarfs of
\cite{Aznar2013}. Since the results of  these calculations is a set of
trajectories of a collection of individual particles some of the usual
expressions must be discretized. Here we explain how we do this.

\subsection{Gravitational waves}
\label{sec:GWR}

We  compute  the  gravitational  wave  emission  in  the  slow-motion,
weak-field quadrupole  approximation \citep{Misner1973}. Specifically,
we follow  closely the procedure outlined  in \cite{Loren2005}. Within
this  approximation the  strain  amplitudes can  be  expressed in  the
following way:
\begin{equation}  
h^{\rm TT}_{jk}(t,\textbf{x})=\frac{G}{c^4 d}\big(A_{+}(t,
\textbf{x})\textbf{e}_{+\, jk}+A_{\times}(t,\textbf{x})
\textbf{e}_{\times\, jk}\big)
\label{pol_decomposition}
\end{equation}
where $d$ is  the distance to the source, and  the polarization tensor
coordinate matrices are defined as:
\begin{eqnarray} 
\textbf{e}_{+\, jk}&=&\frac{1}{\sqrt{2}}[(\textbf{e}_{x})_j
(\textbf{e}_{x})_k-(\textbf{e}_{y})_j(\textbf{e}_{y})_k]\cr
&&\\
\textbf{e}_{\times\, jk}&=&\frac{1}{\sqrt{2}}[(\textbf{e}_{x})_j
(\textbf{e}_{y})_k+(\textbf{e}_{y})_j(\textbf{e}_{x})_k], \nonumber
\end{eqnarray}
and 
\begin{equation}  
A_{+}(t,\textbf{x})= \ddot Q_{xx} - \ddot Q_{yy}, \;\;\; 
A_{\times}(t,\textbf{x})=+2\ddot Q_{xy}
\label{ATT1}
\end{equation}
\noindent for $i=0$, and 
\begin{equation}  
A_{+}(t,\textbf{x})= \ddot Q_{zz} - \ddot Q_{yy}, \;\;\;  
A_{\times}(t,\textbf{x})=-2\ddot Q_{yz}
\label{ATT2}
\end{equation}
for $i=\pi/2$, being $i$ the angle  with respect to the line of sight.
In  these  expressions  $Q$  is  the quadrupole  moment  of  the  mass
distribution.

Since,  as already  mentioned, we  are  post-processing a  set of  SPH
calculations and,  hence, we deal  with an ensemble of  $n$ individual
SPH particles, the double time  derivative of the quadrupole moment is
discretized in the following way:
\begin{eqnarray} 
\ddot Q^{\rm TT}_{jk}&\approx&  P_{ijkl}(\textbf{N})\sum^{n}_{p=1} 
m(p) \lbrack  2\textbf{v}^k(p)\textbf{v}^l(p)\cr
&+&\textbf{x}^k(p)\textbf{a}^l(p)+\textbf{x}^l(p)\textbf{a}^k(p) \rbrack
\label{mom_discretized}
\end{eqnarray}
Where  $m(p)$  is the  mass  of  each SPH  particle,  $\textbf{x}(p)$,
$\textbf{v}(p)$ and  $\textbf{a}(p)$ are, respectively,  its position,
velocity and acceleration, and
\begin{eqnarray} 
P_{ijkl}(\textbf{N})& \equiv & (\delta_{ij}-N_iN_k)(\delta_{jl}-N_jN_l)\cr
&-&\frac{1}{2}(\delta_{ij}-N_iN_j)(\delta_{kl}-N_kN_l)
\label{pol_tensor}
\end{eqnarray}
is  the  transverse-traceless  projection   operator  onto  the  plane
ortogonal to the outgoing wave direction, $\textbf{N}$,

To assess the detectability of  the gravitational waveforms we proceed
as  follows.   For  the  well  defined  elliptical  orbits,  we  first
accumulate the  power of the signal  during one year. We  then compute
the    characteristic   frequencies    and    amplitudes,   and    the
signal-to-noise-ratios  (SNR) according  to \cite{Zanotti2003}.  These
characteristic quantities are given by:
\begin{eqnarray}
h_{\rm c}&=&\left[3\int_0^{\infty}\frac{S_{\rm n}\left(f_{\rm c}\right)}
                {S_n\left(f\right)}\left\langle\left|\tilde{h}\left(f\right)
                \right|^2\right\rangle \,f\, df\right]^{1/2} \\
f_{\rm c}&=&\left[\int_0^{\infty}\frac{\left\langle\left|\tilde{h}\left(f\right)
                \right|^2\right\rangle}{S_n\left(f\right)}\,f\,df\right]
                \left[\int_0^{\infty}\frac{\left\langle\left|\tilde{h}
                \left(f\right)\right|^2\right\rangle}
                {S_{\rm n}\left(f\right)}\,df\right]^{-1} 
\end{eqnarray}
where 
\begin{equation}
\tilde{h}(f)=\int_{-\infty}^{\infty} e^{2\pi ift}\, h(t)\, dt
\end{equation}
is the waveform in the frequency  domain, and $S_{\rm n}$ is the power
spectral  density of  the detector.   After this  the root-mean-square
strain noise:
\begin{equation}
h_{\rm rms}=\sqrt{f S_{\rm n}(f)}
\end{equation}
is computed to obtain the SNR:
\begin{equation}
{\rm SNR}=\frac{h_{\rm c}}{h_{\rm rms}(f_{\rm c})}.
\end{equation} 
For the short-lived signals obtained in lateral collisions, we compute
the SNR as in \cite{Giacomazzo2011}:
\begin{equation}
\left({\rm SNR}\right)^2=4\int_{0}^{\infty}\frac{\left|\tilde{h}
                   \left(f\right)\right|^2}{S_{\rm n}\left(f\right)}
                   df, 
\end{equation}
and  we  compare   the  product  of  the  Fourier   transform  of  the
dimensionless  strains   and  the   square  root  of   the  frequency,
$\tilde{h}(f)  f^{1/2}$, with  the  gravitational-wave detector  noise
curve.   Finally, for  direct  collisions we  only  compute the  total
energy  radiated in  the  form  of gravitational  waves,  since it  is
unlikely that these events could be eventually detected.

\subsection{Light curves}
\label{sec:EM}

The observable  emission of  those events in  which some  $^{56}$Ni is
synthesized  during  the  interaction  is powered  completely  by  its
radioactive  decay  and  that   of  its  daughter  nucleus,  $^{56}$Co
\citep{Colgate}.   Specifically,  the  $^{56}$Ni  synthesized  in  the
explosion decays by  electron capture with a half-life of  6.1 days to
$^{56}$Co, which  in turn decays  through electron capture  (81\%) and
$\beta^+$  decay (19\%)  to stable  $^{56}$Fe with  a half-life  of 77
days.  The  early phase  is dominated by  the down-scattering  and the
release of photons generated as  $\gamma$-rays in the decays, while at
late phases the  optical radiation escapes freely.   The peak radiated
luminosity of SNeIa can be approximated with enough accuracy using the
scaling law of \cite{Arnett1979}.  It  is expected to be comparable to
the instantaneous rate of energy  release by radioactivity at the rise
time \citep{Branch1992}.

To model the light curves  in those interactions where sizable amounts
of   $^{56}$Ni    are   produced,   we   adopt    the   treatment   of
\cite{Kushnir2013}, which provides a  relation between the synthesized
$^{56}$Ni  mass  and  the  late-time  ($\approx$60  days  after  peak)
bolometric  light   curve.   Within   this  approach,   the  late-time
bolometric light  curve is  computed numerically  using a  Monte Carlo
algorithm, which solves the transport of photons, and the injection of
energy by  the $\gamma$-rays produced  by the $^{56}$Ni  and $^{56}$Co
decays.  To this end we first map, using kernel interpolation, our SPH
data  to a  three-dimensional  cartesian velocity  grid,  and then  we
employ the Monte Carlo code of \cite{Kushnir2013}.

\subsection{Thermal neutrinos}
\label{sec:nu}

Since the  densities reached in  all the simulations are  smaller than
$10^{10}$~g~cm$^{-3}$,  the  material  is expected  to  be  completely
transparent to  neutrinos. Moreover, as  in a sizable number  of close
encounters the maximum temperatures are in excess of $10^9$~K, copious
amounts of thermal neutrinos should be emitted.  We thus computed this
emission taking  into account the five  traditional neutrino processes
--- that   is,   electron-positron    annihilation,   plasmon   decay,
photoemission,   neutrino   bremsstrahlung,   as  well   as   neutrino
recombination --- using the prescriptions of \cite{Itoh1996}.

To assess the  possibility of detecting some of  the emitted neutrinos
we  follow   closely  the   prescriptions  of   \cite{neutrinos1}  and
\cite{neutrinos2}.  Specifically, we compute the number of events that
could be eventullay observed in the Super-Kamiokande detector when the
source  is at  a distance  $d=1$~kpc.  We  compute first  the neutrino
spectral flux:
\begin{equation}
 \Phi(E_{\nu},t)=\frac{L}{A
\langle E_{\nu}\rangle^4}\frac{aE_{\nu}^2}
{1+\exp^{bE_{\nu}/\langle E_{\nu}\rangle}}
\end{equation}
where $L$ is the neutrino luminosity  at time $t$, $A=4\pi d^2$ is the
irradiated area,  $\langle E_{\nu}\rangle=3.15137  \, T_{\nu}$  is the
average neutrino  energy at time  $t$, $a\simeq 17.3574$  and $b\simeq
3.15137$.  The values of $L(t)$ and $T_{\nu}(t)$ are obtained from our
SPH  simulations.  Assuming  equipartition of  energy between  all the
emitted neutrino  flavors, the  number of events  detected in  a water
Cherenkov  detector can  be  approximated by  estimating  the rate  of
electron-neutrino  neutrino scatterings.   The cross  section of  this
scattering depends on the threshold  energy of recoil electrons in the
experiment, $T^{\rm th}_{\rm e}$:
\begin{eqnarray}
\sigma(E_{\nu},T^{\rm th}_{\rm e})&=&\frac{\sigma_0}{m_{\rm e}}
\Bigg[\left(g_1^2+g_2^2\right)
\left(T_{\rm e}^{\rm max}-T_{\rm e}^{\rm th}\right)\nonumber \\
&-&\left(g_2^2+g_1 g_2\frac{m_{\rm e}}{2E_{\nu}}\right) 
\left(\frac{T_{\rm e}^{\rm max^{2}}-T_{\rm e}^{\rm th^2}}{E_{\nu}}\right)\nonumber\\
&+&\frac{1}{3}g_2^2
\left(\frac{T_{\rm e}^{\rm max^{3}}-T_{\rm e}^{\rm th^3}}{E_{\nu}^3}\right)\Bigg],
\end{eqnarray}
where $\sigma_0\simeq 88.06\times 10^{-46}\, {\rm cm}^2,$ 
$$T_{\rm e}^{\rm max}=\frac{2 E_{\nu}^2}{m_{\rm e}+2E_{\nu}}$$
is the  maximum kinetic energy  of the  recoil electron at  a neutrino
energy  $E_{\nu}$,  $g_1\simeq  0.73$   and  $g_2  \simeq  0.23$.  The
time-integrated spectra is then
$$F_{\nu}(E_{\nu})=\int \Phi(E_{\nu},t)dt,$$  
and the total number of events is:
$$N=nV \int F_{\nu}(E_{\nu}) \sigma(E_{\nu},T^{\rm th}_{\rm e}) dE_{\nu},$$ 
where $n$ is the number density of electrons in the water tank and $V$
is  its volume.   We consider  the Super-Kamiokande  detector, with  a
fiducial volume of 22,500~tons.

\subsection{Fallback luminosities}
\label{sec:fall}

Another possible observational signature  of these types of collisions
and close encounters  is the emission of high-energy  photons from the
fallback material in the aftermath  of the interaction.  As mentioned,
in  some cases  only one  of the  stars intervening  in the  dynamical
interaction  is  disrupted, and  its  material  goes  to form  a  disk
orbiting the more massive white dwarf.   Most of the material of these
disks has  circularized orbits. However,  it turns out that  a general
feature of these interactions is that some of the SPH particles of the
disk  have highly  eccentric orbits.   After some  time this  material
interacts with the rest of  the disk.  As shown by \cite{Rosswog2007},
the relevant timescale for these particles is not the viscous one, but
instead their evolution is set  by the distribution of eccentricities.
To compute  how these particles  interact with the newly  formed disk,
and which  are the  corresponding fallback accretion  luminosities, we
follow closely the approach described in \cite{Rosswog2007}.  That is,
we assume  that the  kinetic energy of  these particles  is dissipated
within the radius of the disk.

\section{Results}
\label{sec:results}

In this section we present  the possible observational consequences of
the  interactions. The  section  is organized  as  follows.  We  first
describe  the resulting  gravitational waveforms.   We do  so for  the
three  possible  outcomes  of  the  interactions.   This  is  done  in
Sects.~\ref{sec:GWR:EO}     to    \ref{sec:GWR:DC}.      It    follows
Sect.~\ref{sec:results:LCs}  where we  present  the  light curves  for
those  interactions  in  which   a  detonation,  with  the  subsequent
disruption    of   one    of   the    stars,   occurs,    whereas   in
Sect.~\ref{sec:results:nu} we  discuss the neutrino emission  of these
interactions.   Finally, in  Sect.~\ref{sec:results:Xray} we  describe
the  fallback X-ray  luminosities for  those interactions  in which  a
debris region around a central white dwarf is formed.

\subsection{Gravitational wave radiation}
\label{sec:results:GWR}

\subsubsection{Eccentric orbits}
\label{sec:GWR:EO}

\begin{figure}
   \resizebox{\hsize}{!}
   {\includegraphics[width=\columnwidth]{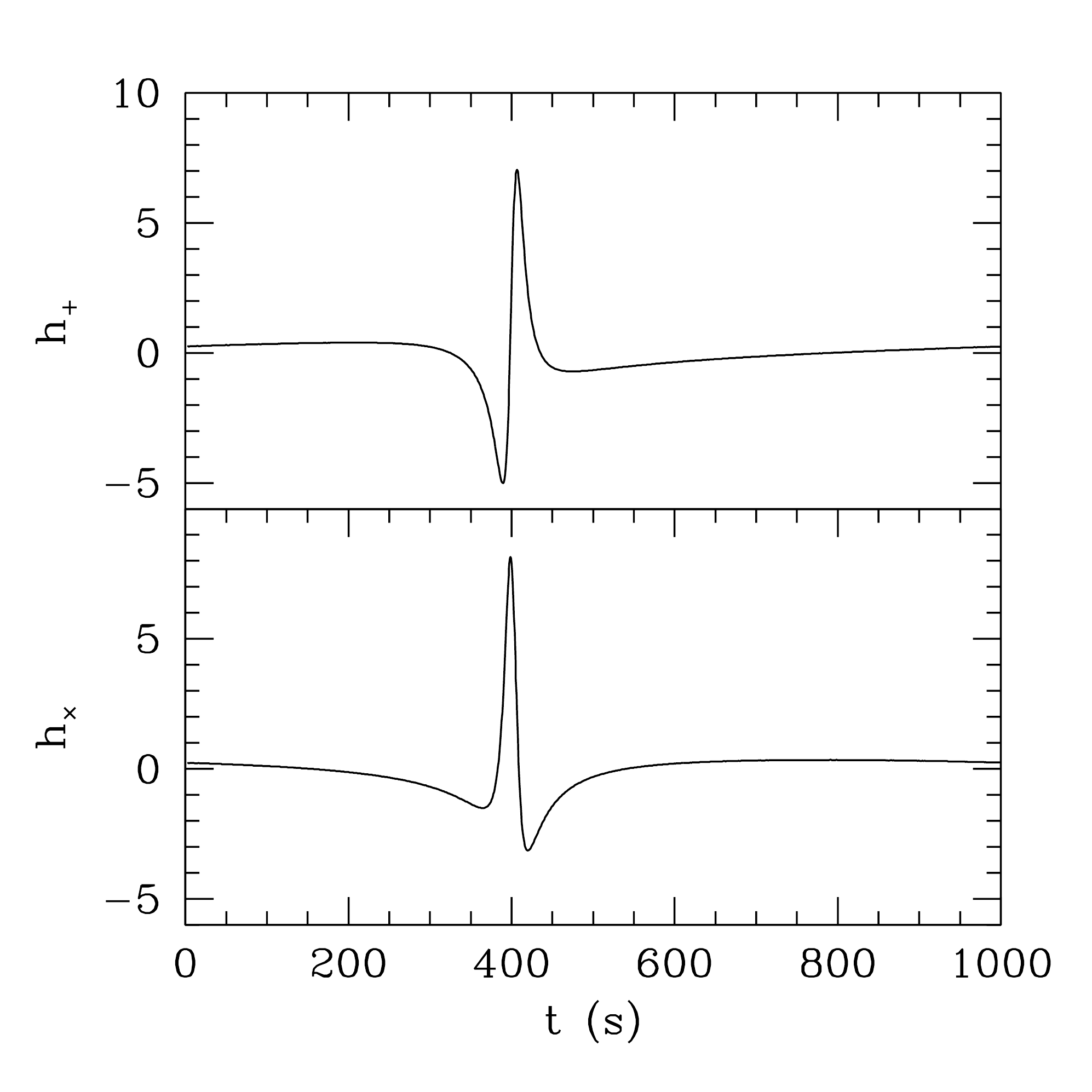}}
   \caption{Gravitational waveforms for a  close encounter in which an
     eccentric binary is  formed. For this particular  case both white
     dwarfs have equal masses,  $M=0.8\, M_{\sun}$, whilst the adopted
     initial conditions are  $v_{\rm ini}=200$~km~s$^{-1}$ and $\Delta
     y=0.4\, R_{\sun}$.  We only show  the waveforms $h_+$ (top panel)
     and $h_\times$  (bottom panel)  for an  inclination $i=0^{\circ}$
     and a  distance of 10~kpc, in  units of $10^{-22}$. See  text for
     details.}
\label{fig:eo:gw}
\end{figure}

As mentioned, the  most complete and comprehensive study  of the close
encounters  of two  white dwarfs  is that  of \cite{Aznar2013},  which
covers a considerably  wide range of masses  and chemical compositions
of the interacting white dwarfs, as  well as of initial conditions. In
particular, they characterize  the initial conditions in  terms of the
initial  relative velocity  ($v_{\rm ini}$)  and the  initial distance
perpendicular to the relative velocity  ($\Delta y$) at a sufficiently
large  distance so  that  the  two interacting  white  dwarfs are  not
affected  by tidal  forces.  Here  we will  compute the  gravitational
waveforms of  the interactions computed  by these authors and  we will
refer to them employing this notation.

\begin{figure}
   \resizebox{\hsize}{!}
   {\includegraphics[width=\columnwidth]{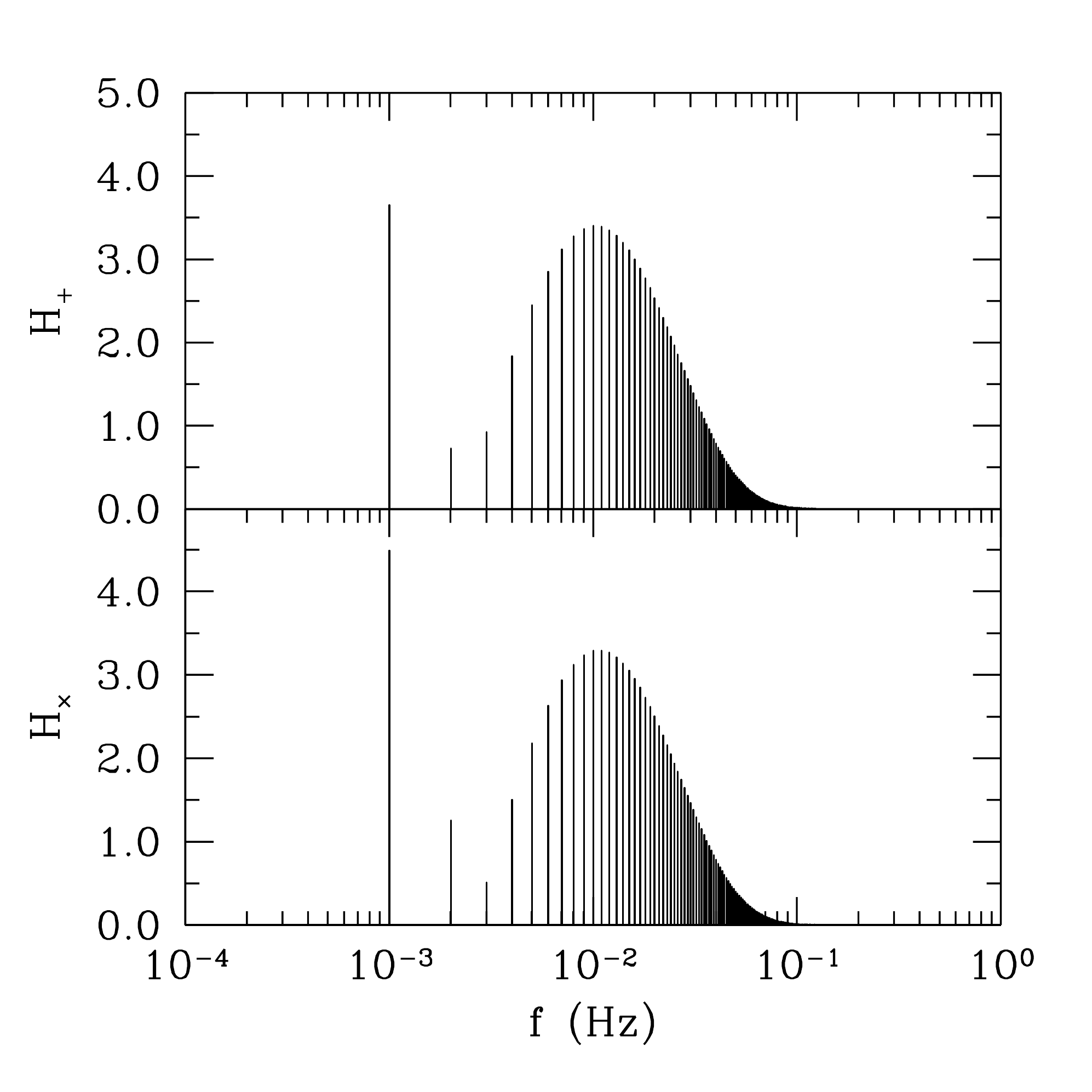}}
   \caption{Fourier  transforms  of  the  gravitational  waveforms  of
     Fig.~\ref{fig:eo:gw}, in  units of $10^{-23}$.}
\label{fig:eo:ft}
\end{figure}

\begin{table}
\caption{Signal-to-noise  ratios of  the  gravitational  waves of  the
  interactions resulting in  eccentric orbits, for the  case of eLISA,
  adopting $i=0^\circ$ and a distance of 10~kpc.}
\label{tab:eo:SNR}
\begin{center}
\small
\begin{tabular}{ccccc}
\hline 
\hline 
\noalign{\smallskip}
 Run & $M_1+M_2$  & $e$ & $f$ & SNR \\
     & ($M_{\sun}$) &  & (Hz) & \\[0.6ex]
\noalign{\smallskip}
\hline 
\hline
\multicolumn{5}{l}{$v_{\rm ini} = 150 \;\text{km/s} \quad \quad \Delta y = 0.4 \;R_{\sun}$} \\
\hline
\noalign{\smallskip}
55	&	0.2+0.4		&	0.736	&	5.54$\times 10^{-4}$	&	1.90	\\
56	&	0.4+0.4		&	0.799	&	6.91$\times 10^{-4}$	&	5.25	\\
\hline						
\multicolumn{5}{l}{$v_{\rm	ini}	= 200 \;\text{km/s} \quad \quad	\Delta y = 0.3 \;R_{\sun}$} \\
\hline							
\noalign{\smallskip}									
59	&	0.2+0.4		&	0.695	&	6.64$\times 10^{-4}$	&	2.00	\\
60	&	0.4+0.4		&	0.762	&	8.64$\times 10^{-4}$	&	5.87    \\
\hline
\multicolumn{5}{l}{$v_{\rm ini} = 200 \;\text{km/s} \quad \quad \Delta y	= 0.4 \;R_{\sun}$} \\
\hline								
\noalign{\smallskip}										
25	&	0.8+0.6		&	0.796	&	9.11$\times 10^{-4}$	&	16.41	\\
26	&	0.8+0.8		&	0.820	&	1.00$\times 10^{-3}$	&	23.42	\\
27	&	1.0+0.8		&	0.840	&	1.09$\times 10^{-3}$	&	30.48	\\
28	&	1.2+0.8		&	0.855	&	1.16$\times 10^{-3}$	&	37.38	\\
63	&	0.2+0.4		&	0.564	&	4.25$\times 10^{-4}$	&	0.48	\\
64	&	0.4+0.4		&	0.657	&	5.74$\times 10^{-4}$	&	2.49	\\
65	&	0.8+0.4		&	0.764	&	8.12$\times 10^{-4}$	&	9.70	\\
\hline							
\multicolumn{5}{l}{$v_{\rm ini} = 300 \;\text{km/s} \quad \quad \Delta y	= 0.3 \;R_{\sun}$} \\
\hline
\noalign{\smallskip}										
29	&	0.8+0.6		&	0.705	&	1.03$\times 10^{-3}$	&	15.11	\\
30	&	0.8+0.8		&	0.736	&	1.17$\times 10^{-3}$	&	22.90	\\
31	&	1.0+0.8		&	0.762	&	1.30$\times 10^{-3}$	&	31.08	\\
32	&	1.2+0.8		&	0.783	&	1.41$\times 10^{-3}$	&	39.33	\\
67	&	0.4+0.4		&	0.571	&	5.21$\times 10^{-4}$	&	1.38	\\
68	&	0.8+0.4		&	0.666	&	8.82$\times 10^{-4}$	&	8.11	\\
69	&	1.2+0.4		&	0.737	&	1.17$\times 10^{-3}$	&	17.22	\\
\hline
\multicolumn{5}{l}{$v_{\rm ini} = 300 \;\text{km/s} \quad \quad \Delta y	= 0.4 \;R_{\sun}$} \\										
\hline
\noalign{\smallskip}
33	&	0.8+0.6		&	0.576	&	6.64$\times 10^{-4}$	&	6.04	\\
34	&	0.8+0.8		&	0.620	&	7.67$\times 10^{-4}$	&	10.71	\\
35	&	1.0+0.8		&	0.657	&	8.62$\times 10^{-4}$	&	16.90	\\
36	&	1.2+0.8		&	0.688	&	9.49$\times 10^{-4}$	&	23.77	\\
70	&	0.8+0.4		&	0.525	&	5.51$\times 10^{-4}$	&	2.99	\\
71	&	1.2+0.4		&	0.621	&	7.67$\times 10^{-4}$	&	8.05	\\
\noalign{\smallskip}
\hline
\hline
\end{tabular}
\end{center}
\end{table}

As  previously explained,  some of  the close  encounters computed  by
\cite{Aznar2013}  result in  the  formation of  a  binary system  that
survives the  interaction and no  mass transfer between the  two stars
occurs.  In these cases the eccentricity of the orbit is always large,
$e>0.5$.   Fig.~\ref{fig:eo:gw}  shows   the  resulting  gravitational
dimensionless strains of  a typical case, in units  of $10^{-22}$, for
an  edge-on  orbit.   The   two  polarizations  of  the  gravitational
radiation, $h_+$ and  $h_\times$, are shown, respectively,  in the top
and bottom panels,  for a full period of the  orbit. The corresponding
dimensionless Fourier  transforms, $H(f)=\tilde{h}/T$  where $T=1$~yr,
of   these    gravitational   wave    patterns   are    displayed   in
Fig.~\ref{fig:eo:ft}.  As can be seen, the gravitational wave emission
presents  a  single,   narrow,  large  pulse,  which   occurs  at  the
periastron, whereas  the corresponding  Fourier transforms,  $H_+$ and
$H_\times$,  present a  peak  at the  orbital  frequency, and  several
Fourier components  at larger  frequencies.  This  was expected,  as a
binary   system  in   a  circular   orbit  produces   a  monochromatic
gravitational  waveform  at twice  the  orbital  frequency, while  for
eccentric   orbits    it   is    found   ---   see,    for   instance,
\cite{Wahlquist1987} --- that the binary  system radiates power at the
fundamental mode  and at several  harmonics of the  orbital frequency.
The reason for this is clear, since it is at closest approach when the
accelerations are  larger, thus resulting  in an enhanced  emission of
gravitational  radiation. Moreover,  it  turns out  that  for a  fixed
energy,  the larger  the  eccentricity, the  smaller  the distance  at
closest approach, and therefore the larger the radiated power.

\begin{figure}
   \resizebox{\hsize}{!}
   {\includegraphics[width=\columnwidth]{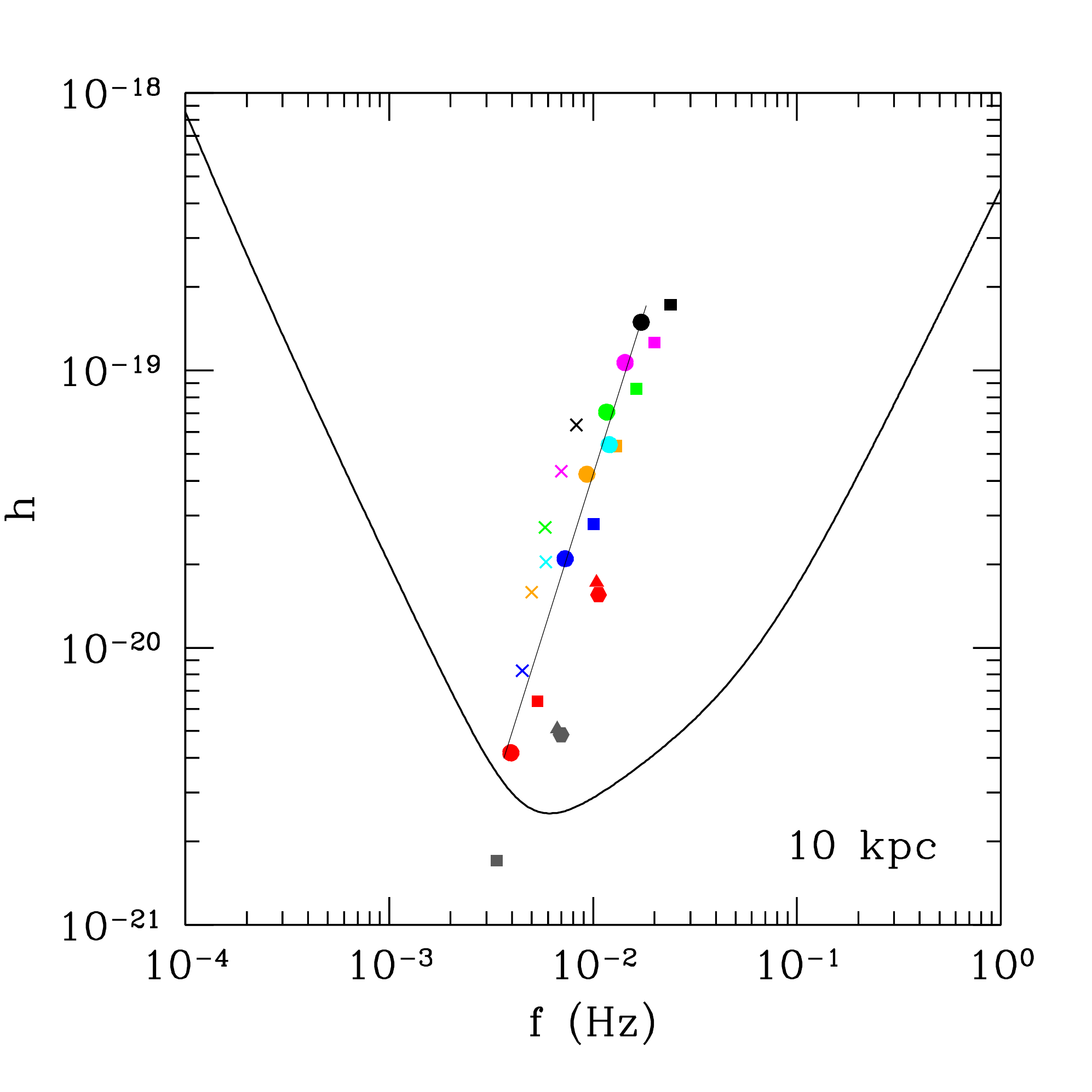}}
   \caption{A comparison  of the  signal produced  by the  close white
     dwarf binary systems  studied here, when a distance  of 10~kpc is
     adopted, with the  spectral distribution of noise of  eLISA for a
     one  year integration  period, and  for a  null inclination.   We
     assess the detectability of  the $h_+$ dimensionless strain.  The
     different colors denote different masses of the interacting white
     dwarfs,  whereas the  different symbols  are used  to distinguish
     between different initial conditions. Specifically, black symbols
     correspond to  a $1.2\,  M_{\sun}+0.8\, M_{\sun}$  binary system,
     magenta ones  to a $1.0\, M_{\sun}+0.8\,M_{\sun}$,  green symbols
     to  a $0.8\,  M_{\sun}+0.8\,  M_{\sun}$, cyan  ones  to a  $1.2\,
     M_{\sun}+0.4\,   M_{\sun}$,   orange    symbols   to   a   $0.8\,
     M_{\sun}+0.6\, M_{\sun}$,  blue ones  to a  $0.8\, M_{\sun}+0.4\,
     M_{\sun}$, red symbols to  a $0.4\, M_{\sun}+0.4\, M_{\sun}$, and
     grey ones are  used for a $0.4\,  M_{\sun}+0.2\, M_{\sun}$ binary
     system. On  the other  hand, hexagons  are used  for the  case in
     which  $v_{\rm  ini}=150$~km/s  and  $\Delta  y=0.4\,  R_{\sun}$,
     triangles   for  $v_{\rm   ini}=200$~km/s  and   $\Delta  y=0.3\,
     R_{\sun}$,  squares   for  $v_{\rm  ini}=200$~km/s   and  $\Delta
     y=0.4\,R_{\sun}$, circles for  $v_{\rm ini}=300$~km/s and $\Delta
     y=0.3\, R_{\sun}$, crosses for $v_{\rm ini}=300$~km/s and $\Delta
     y=0.4\, R_{\sun}$. See  the on-line edition of the  journal for a
     color version of this figure.}
\label{fig:eo:elisa}
\end{figure}

In  Table~\ref{tab:eo:SNR}   we  list   for  each   of  the   runs  of
\cite{Aznar2013} which result in the  formation of an eccentric binary
system the masses of the interacting white dwarfs, the eccentricity of
the  orbit,   the  frequency   of  the   fundamental  mode,   and  the
signal-to-noise  ratios  for $h_+$  for  the  eLISA mission,  adopting
$i=0^\circ$ and  a distance of  10~kpc. Also  listed, for the  sake of
completeness,  is  the   run  number  (first  column)   as  quoted  in
\cite{Aznar2013}. As can be seen, eLISA  will be able to detect almost
all these  systems with sufficiently  large SNRs. Only run  number 63,
which corresponds to the binary system with an orbit with the smallest
eccentricity  ($e=0.564$)  and for  which  the  two interacting  white
dwarfs have  the smallest  masses --- namely,  $M_1=0.4\,M_{\sun}$ and
$M_2=0.2\,M_{\sun}$  --- will  not  be detectable.   The  rest of  the
interactions have, in general, large SNRs.

\begin{figure}
   \resizebox{\hsize}{!}
   {\includegraphics[width=\columnwidth]{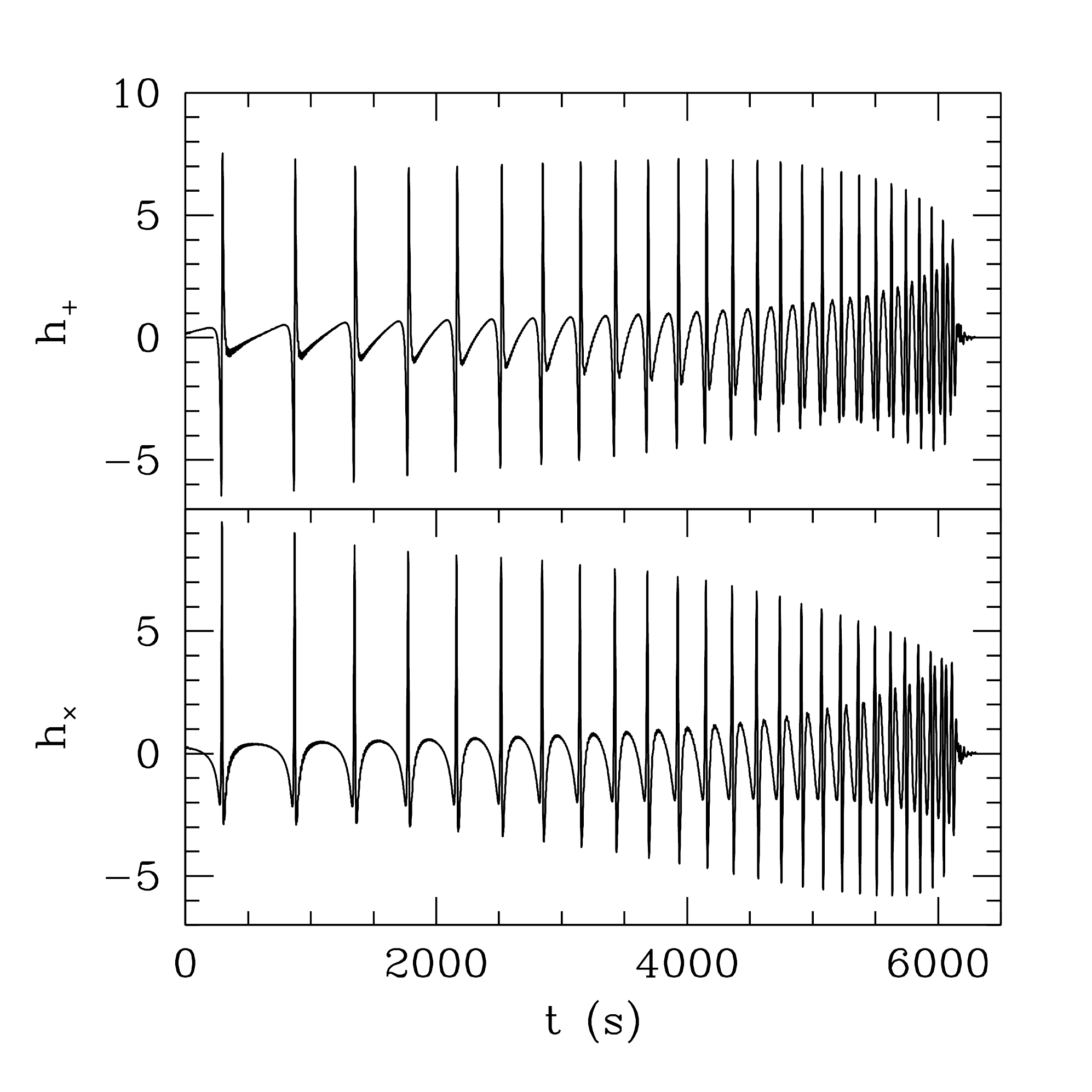}}
   \caption{Gravitational waveforms for a close encounter in which the
     outcome  of the  interaction  is a  lateral  collision. For  this
     particular case two  white dwarfs of masses  $0.8\, M_{\sun}$ and
     $0.6\,  M_{\sun}$,  respectily,  collide.   The  adopted  initial
     conditions of this  interaction are $v_{\rm ini}=200$~km~s$^{-1}$
     and $\Delta y=0.3\, R_{\sun}$. As in Fig.~\ref{fig:eo:gw} we only
     show  the  waveforms $h_+$  (top  panel)  and $h_\times$  (bottom
     panel) for an inclination $i=0^{\circ}$ and a distance of 10~kpc.
     Again,   the  units   of  of   the  dimensionless   stresses  are
     $10^{-22}$.}
\label{fig:LC:gw}
\end{figure}

In  Fig.~\ref{fig:eo:elisa} we  display the  characteristic amplitudes
and frequencies of the $h_+$  dimensionless strain for $i=0^\circ$ and
we compare  them with the  sensitivity curve of eLISA,  computed using
the  analytical  approximation  of  \cite{Amaro2013}.   As  previously
noted,  only   run  number  63   lies  below  the   eLISA  sensitivity
limit. Moreover,  in this  figure it  can be clearly  seen that  for a
fixed initial  condition ---  that is  for a fixed  pair of  values of
$v_{\rm ini}$ and $\Delta y$  --- the characteristic amplitude as well
as the frequency decreases with the total mass.  The mass ratio of the
interacting white dwarfs, $q={M_2}/{M_1}$, also plays a role. This can
be seen when simulations  30 and 69, or 34 and  71, are compared. Both
pairs of simulations have the  same initial conditions and total mass,
but  different values  of  $q$.  As  can  be seen,  the  value of  the
characteristic amplitude in runs 69 and  71 --- which have $q=1/3$ ---
is smaller, while the characteristic frequencies are larger than those
of runs 30  and 34, respectively --- which have  $q=1$. Finally, it is
interesting to realize that for a  given set of initial conditions the
runs of  different masses lie  approximately on straight  lines, which
are not  all shown  for the  sake of clarity.  In particular,  we only
show,  using  a  thin  solid  line, the  case  in  which  the  initial
conditions are $v_{\rm ini}=300$~km/s and $\Delta y=0.3\, R_{\sun}$.

\subsubsection{Lateral collisions}
\label{sec:GWR:LC}

In lateral collisions the less massive white dwarf is tidally deformed
by the more massive star at closest approach to such an extent that in
the end  some of its  material is  accreted by the  massive companion.
This occurs in several mass  tranfer episodes, and the resulting final
configuration consists  of a  central compact  object surrounded  by a
hot, rapidly rotating corona, and an external region where some of the
debris  produced during  the dynamical  interaction can  be found.   A
typical example of  the gravitational wave pattern  resulting in these
cases  is  shown in  Fig.~\ref{fig:LC:gw},  which  corresponds to  run
number 21 of \cite{Aznar2013}. This specific simulation corresponds to
the  dynamical interaction  of  two white  dwarfs  with masses  $0.8\,
M_{\sun}$ and $0.6\, M_{\sun}$, whereas the adopted initial conditions
were    $v_{\rm    ini}=200$~km~s$^{-1}$     and    $\Delta    y=0.3\,
R_{\sun}$. Again, in this figure we only show the waveforms $h_+$ (top
panel) and $h_\times$ (bottom  panel) for an inclination $i=0^{\circ}$
and a distance of 10~kpc, in units of $10^{-22}$.  As can be seen, the
time  evolution of  the  dimensionless strains  presents  a series  of
peaks. For this specific case each one of these peaks corresponds to a
mass transfer  episode, which occurs  short after the  passage through
the periastron. Nevertheless, it is to be noted that, depending on the
masses  of the  stars  and  on the  initial  conditions  of the  close
encounter, the eccentricity  of the orbit and the  distance at closest
approach may  be quite  different for  the several  lateral collisions
studied here.  Hence,  the number of periastron passages  shows a wide
range of variation.   However, a general feature in all  cases is that
the emission of gravitational waves  is largest for the first passages
through  the periastron  and  the  amplitude of  the  narrow peaks  of
gravitational wave  radiation decreases in subsequent  passages.  Note
as  well that  the time  difference  between successive  peaks of  the
gravitational wave pattern also decreases  as time passes by. All this
is a  consequence of  the fact  that after  every passage  through the
periastron the  orbit is  slightly modified,  either because  the less
massive white dwarf is substantially deformed by tidal forces in those
cases in which during the  first passages through the periastron there
is no  mass transfer  from the  less massive white  dwarf to  the more
massive  one, or  because mass  transfer has  happened, and  the orbit
becomes  less eccentric.   In the  former case,  after a  few passages
through  the  periastron  mass  transfer occurs,  and  the  subsequent
evolution is similar to that of lateral collisions in which there is a
mass  transfer episode  during the  first closest  approach.  Finally,
after several mass  transfers, the binding energy of  the less massive
star  becomes positive  and  it  is totally  destroyed,  leading to  a
merger.  This  causes the  sudden decrease  of the  gravitational wave
emission,  although some  oscillations  of the  remnant still  radiate
gravitational waves  --- in  a way  very much similar  to that  of the
ring-down phase found in mergers of two compact objects --- but with a
significantly smaller amplitude than the previous ones.

\begin{figure}
   \resizebox{\hsize}{!}
   {\includegraphics[width=\columnwidth]{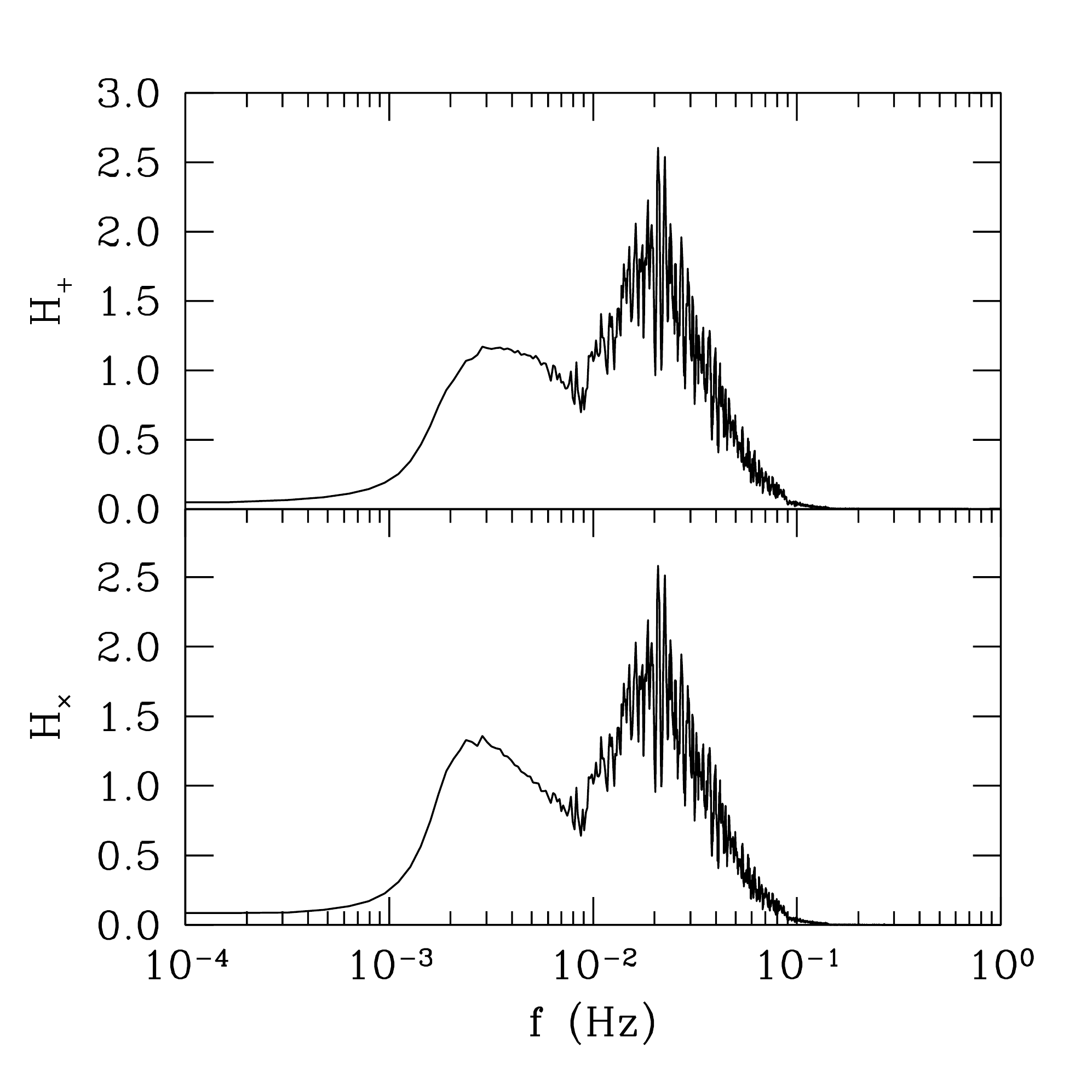}}
   \caption{Fourier  transforms  of  the  gravitational  waveforms  of
     Fig.~\ref{fig:LC:gw}, in units of $10^{-23}$.}
\label{fig:LC:ft}
\end{figure}

The dimensionless Fourier transform  of the gravitational wave pattern
shown  in Fig.~\ref{fig:LC:gw}  is displayed  in Fig.~\ref{fig:LC:ft}.
The   only  difference   with  those   Fourier  transforms   shown  in
Fig.~\ref{fig:eo:ft} is that in this case to compute the dimensionless
Fourier transform  we adopt the  duration of the merger.   In contrast
with  the  Fourier  transform   of  the  gravitational  wave  emission
presented  in  Fig.~\ref{fig:eo:ft},  which is  discrete  because  the
signal  is  periodic,  now  the emission  of  gravitational  waves  is
characterized by  a continuous spectrum,  and presents two  peaks. The
first  one is  a broad,  smooth peak  at around  $f=3$~mHz, while  the
second one,  which is of  larger amplitude  but more noisy,  occurs at
larger   frequencies,  $\sim   20$~mHz.    Since   in  our   numerical
configuration all the lateral collisions have orbits which can be well
approximated until  the passage  through the periastron  by elliptical
ones, the first of these peaks  corresponds to the fundamental mode of
the eccentric  orbit, while the second  one is due to  the presence of
higher order  harmonics, as it  occurs for the pure  elliptical orbits
previously   analyzed  in   Sect.~\ref{sec:GWR:EO},  being   the  most
significant difference the continuous shift  in frequency due to orbit
circularization, which noticeably broadens the fundamental mode.

\begin{table*}
\caption{Properties of lateral collisions.}
\label{tab:LC:SNR}
\begin{center}
\small
\begin{tabular}{cccccccc}
\hline 
\hline 
\noalign{\smallskip}
\noalign{\smallskip}
 Run & $M_1+M_2$      & $N$ &  $M_{\rm Ni}$   & $E_{\rm GW}$ &  $E_{\nu}$ &  \multicolumn{2}{c}{SNR} \\
\cline{7-8}             
     & ($M_{\sun}$)   &     & ($M_{\sun}$)    & \multicolumn{2}{c}{(erg)} &    eLISA      &    ALIA  \\
\noalign{\smallskip}
\hline
\hline
\multicolumn{8}{l}{$v_{\rm ini} = 75 \;\text{km/s} \quad \quad \Delta y = 0.4 \;R_{\sun}$} \\
\hline
\noalign{\smallskip}
39 & 0.4+0.2 & 2 & 0                     & 2.78$\times 10^{38}$ & 3.03$\times 10^{42}$ & 0.02 &  1.60 \\
\hline
\multicolumn{8}{l}{$v_{\rm ini} = 100 \;\text{km/s} \quad \quad \Delta y = 0.3 \;R_{\sun}$} \\
\hline
\noalign{\smallskip}
43 & 0.4+0.2 & 2 & 4.97$\times 10^{-14}$ & 2.69$\times 10^{38}$ & 3.23$\times 10^{42}$ & 0.02 &  1.61 \\
\hline
\multicolumn{8}{l}{$v_{\rm ini} = 100 \;\text{km/s} \quad \quad \Delta y = 0.4 \;R_{\sun}$} \\
\hline
\noalign{\smallskip}
9  & 0.6+0.8 & 2 & 0                     & 1.32$\times 10^{41}$ & 3.64$\times 10^{43}$ & 0.11 & 12.24 \\
10 & 0.8+0.8 & 2 & 4.52$\times 10^{-10}$ & 6.37$\times 10^{41}$ & 2.25$\times 10^{44}$ & 0.15 & 18.03 \\
11 & 1.0+0.8 & 2 & 7.94$\times 10^{-7}$  & 1.16$\times 10^{42}$ & 1.28$\times 10^{45}$ & 0.16 & 18.74 \\
12 & 1.2+0.8 & 2 & 1.09$\times 10^{-5}$  & 2.10$\times 10^{42}$ & 2.43$\times 10^{45}$ & 0.16 & 18.73 \\
47 & 0.4+0.2 & 2 & 0                     & 1.87$\times 10^{38}$ & 3.58$\times 10^{35}$ & 0.02 &  1.58 \\
48 & 0.4+0.4 & 2 & 2.76$\times 10^{-11}$ & 4.36$\times 10^{39}$ & 3.24$\times 10^{43}$ & 0.07 &  5.66 \\
\hline
\multicolumn{8}{l}{$v_{\rm ini} = 150 \;\text{km/s} \quad \quad \Delta y = 0.3 \;R_{\sun}$} \\
\hline
\noalign{\smallskip}
13 & 0.6+0.8 & 2 & 0                     & 1.18$\times 10^{41}$ & 1.82$\times 10^{41}$ & 0.13 & 13.47 \\
14 & 0.8+0.8 & 2 & 0                     & 4.45$\times 10^{41}$ & 4.36$\times 10^{43}$ & 0.17 & 19.38 \\
15 & 1.0+0.8 & 2 & 1.56$\times 10^{-13}$ & 8.58$\times 10^{41}$ & 8.82$\times 10^{43}$ & 0.17 & 20.16 \\
16 & 1.2+0.8 & 2 & 9.70$\times 10^{-10}$ & 1.32$\times 10^{42}$ & 6.12$\times 10^{44}$ & 0.14 & 15.75 \\
51 & 0.4+0.2 & 2 & 0                     & 1.15$\times 10^{38}$ & 1.70$\times 10^{35}$ & 0.03 &  1.38 \\
52 & 0.4+0.4 & 2 & 0                     & 4.04$\times 10^{39}$ & 1.16$\times 10^{43}$ & 0.07 &  6.33 \\
\hline
\multicolumn{8}{l}{$v_{\rm ini} = 150 \;\text{km/s} \quad \quad \Delta y = 0.4 \;R_{\sun}$} \\
\hline
\noalign{\smallskip}
17 & 0.6+0.8 & 6 & 0                     & 8.77$\times 10^{40}$ & 2.97$\times 10^{39}$ & 0.28 & 24.25 \\
18 & 0.8+0.8 & 4 & 0                     & 4.57$\times 10^{41}$ & 3.74$\times 10^{42}$ & 0.29 & 31.33 \\
19 & 1.0+0.8 & 5 & 0			 & 5.61$\times 10^{41}$ & 1.56$\times 10^{41}$ & 0.36 & 36.83 \\
20 & 1.2+0.8 & 3 & 0                     & 6.26$\times 10^{41}$ & 2.44$\times 10^{43}$ & 0.31 & 32.34 \\
57 & 0.8+0.4 & 3 & 6.98$\times 10^{-14}$ & 7.01$\times 10^{39}$ & 1.83$\times 10^{42}$ & 0.12 &  8.56 \\
58 & 1.2+0.4 & 2 & 5.61$\times 10^{-10}$ & 2.55$\times 10^{40}$ & 1.63$\times 10^{43}$ & 0.11 &  9.70 \\
\hline
\multicolumn{8}{l}{$v_{\rm ini} = 200 \;\text{km/s} \quad \quad \Delta y = 0.3 \;R_{\sun}$} \\
\hline
\noalign{\smallskip}
21 & 0.6+0.8 & 27 & 0                    & 2.17$\times 10^{41}$ & 9.45$\times 10^{39}$ & 0.57 & 45.83 \\
22 & 0.8+0.8 & 20 & 0                    & 1.07$\times 10^{42}$ & 6.60$\times 10^{39}$ & 0.64 & 63.28 \\
23 & 1.0+0.8 & 5  & 0                    & 5.69$\times 10^{41}$ & 2.46$\times 10^{41}$ & 0.37 & 37.69 \\
24 & 1.2+0.8 & 3  & 0                    & 5.82$\times 10^{41}$ & 3.84$\times 10^{44}$ & 0.31 & 32.10 \\
61 & 0.8+0.4 & 3  & 0                    & 7.03$\times 10^{39}$ & 3.85$\times 10^{41}$ & 0.12 &  8.70 \\
62 & 1.2+0.4 & 2  & 3.27$\times 10^{-9}$ & 2.03$\times 10^{40}$ & 2.36$\times 10^{43}$ & 0.11 &  9.67 \\
\hline
\multicolumn{8}{l}{$v_{\rm ini} = 200 \;\text{km/s} \quad \quad \Delta y = 0.4 \;R_{\sun}$}  \\
\hline
\noalign{\smallskip}
66 & 1.2+0.4 & 8  & 7.30$\times 10^{-12}$ & 2.46$\times 10^{40}$ & 5.89$\times 10^{42}$ & 0.25 & 15.50\\
\noalign{\smallskip}
\hline
\hline
\end{tabular}
\end{center}
\end{table*}

\begin{figure}
   \resizebox{\hsize}{!}
   {\includegraphics[width=\columnwidth]{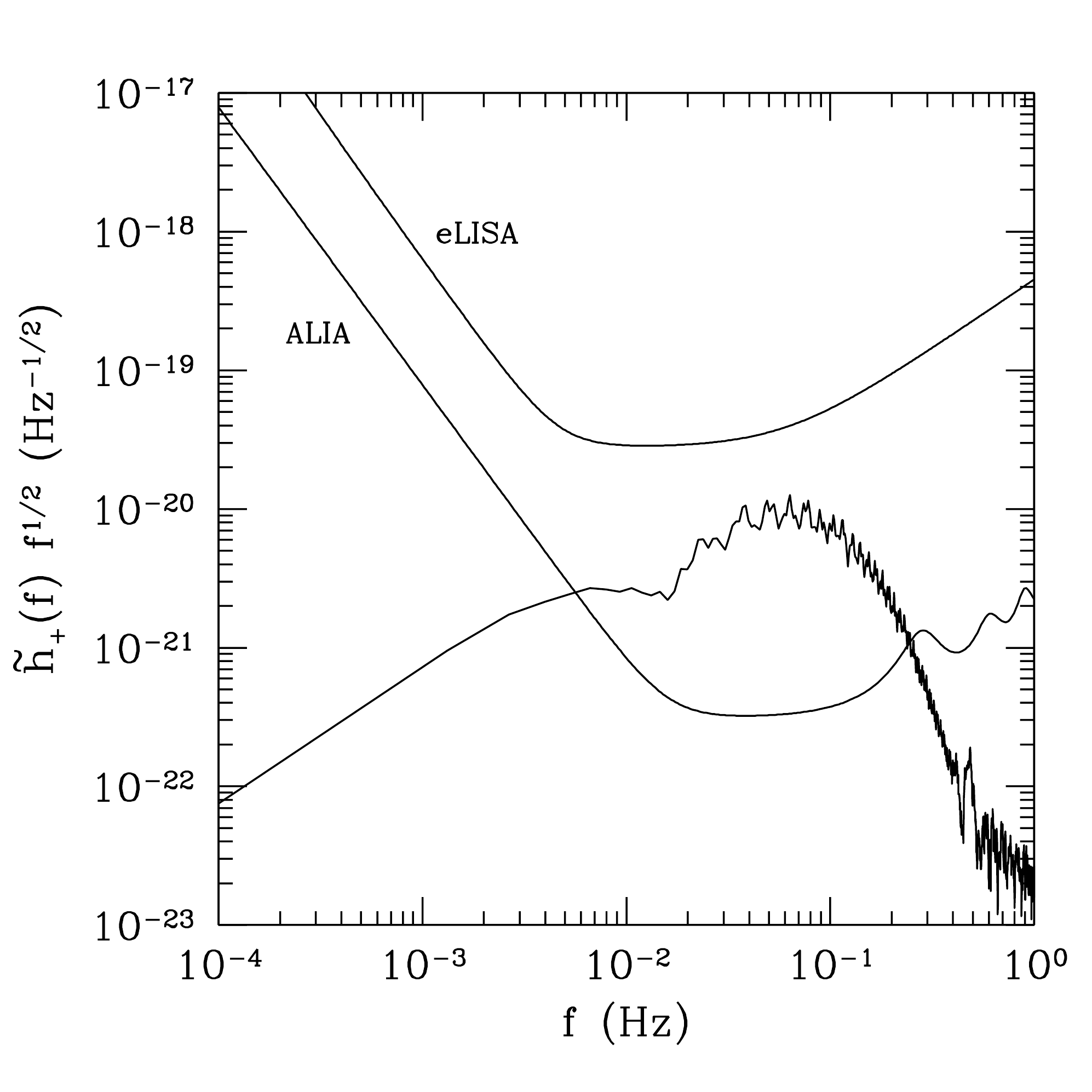}}
   \caption{Fourier spectra of the gravitational waveform of a typical
     lateral collision  --- run  number 20,  which involves  two white
     dwarfs  of   masses  $0.8\,   M_{\sun}$  and   $1.2\,  M_{\sun}$,
     respectively  --- in  units that  facilitate the  comparison with
     gravitational-wave detector  noise curves.  The noise  spectra of
     eLISA and ALIA are also shown for comparison.}
\label{fig:LC:elisa}
\end{figure}

In Fig.~\ref{fig:LC:elisa}  we compare the amplitude  of the simulated
waveform of run number 20 (a  typical lateral collision) as a function
of the frequency to the strain sensitivity of two detectors, eLISA and
ALIA. As can be seen, eLISA will  not be able to detect this dynamical
interaction at a distance of 10~kpc.  However, ALIA \citep{ALIA} --- a
proposed space-borne gravitational-wave  detector --- would eventually
be  able to  detect  most  of these  gravitational  signals.  This  is
quantified  in   table~\ref{tab:LC:SNR},  where   we  list   for  each
simulation, the masses of the  interacting white dwarfs, the number of
mass transfer episodes  ($N$), the mass of nickel  synthesized (if the
mass of $^{56}$Ni is zero means that the interaction failed to produce
a detonation, or  that the detonation conditions were  reached in very
small region of the shocked material, for very short periods of time),
the energy radiated in the form of gravitational waves, and the energy
carried away by  thermal neutrinos.  Finally, in the  last two columns
we list the  SNR of $h_+$ for the same  conditions used previously. In
general, for a fixed white dwarf  pair, the less eccentric orbits with
larger periastron  distances the  larger the  number of  mass transfer
episodes, and the  smaller the maximum amplitude  of the gravitational
wave signal.   This is an  expected result,  as a less  violent merger
episode    lasting   for    longer    times    results   in    smaller
accelerations. Nevertheless,  none of  the lateral  collisions studied
here has  chances to be detected  by eLISA, given that  the respective
SNRs are always relatively small,  a quite unfortunate situation. More
sophisticated and sensitive detectors ---  like ALIA, for instance ---
would, however, be able to detect such events.

Table~\ref{tab:LC:SNR} deserves further comments. In particular, it is
worth  mentioning  that the  total  energy  released  in the  form  of
gravitational waves  increases as  the total mass  of the  white dwarf
pair increases,  as it  should be  expected.  More  interestingly, the
mass ratio  of the  colliding white dwarfs,  $q=M_2/M_1$ also  plays a
role.  Specifically, we find that for  a fixed total mass, the smaller
the  value of  $q$,  the  smaller the  strength  of the  gravitational
signals.    Additionally,  since   the   total   energy  radiated   as
gravitational waves  not only depends  on the masses of  the colliding
white dwarfs, but  also on the duration of  the dynamical interaction,
we  find that,  for fixed  initial conditions,  lateral collisions  in
which the total mass of the  system is large can release small amounts
of energy in the form of  gravitational waves, depending on the number
of mass transfer  episodes.  Actually, two of the cases  for which the
SNR  is largest  --- simulations  number 21  and 22,  which involve  a
$M_1=0.6\,  M_{\sun}$  and  a  $M_2=\, 0.8  M_{\sun}$  and  two  $0.8,
M_{\sun}$ white  dwarfs ---  clearly correspond to  those interactions
which have  longest durations,  and more  mass transfer  episodes, and
nevertheless  the  masses of  the  interacting  white dwarfs  are  not
excesively  large.   Note,   however,  that  in  run   number  20  the
gravitational energy released is larger than in simulation 21, but the
SNR is smaller. This occurs because  run 21 peaks at frequencies where
eLISA and ALIA will be most sensitive.

\subsubsection{Direct collisions}
\label{sec:GWR:DC}

\begin{figure}
   \resizebox{\hsize}{!}
   {\includegraphics[width=\columnwidth]{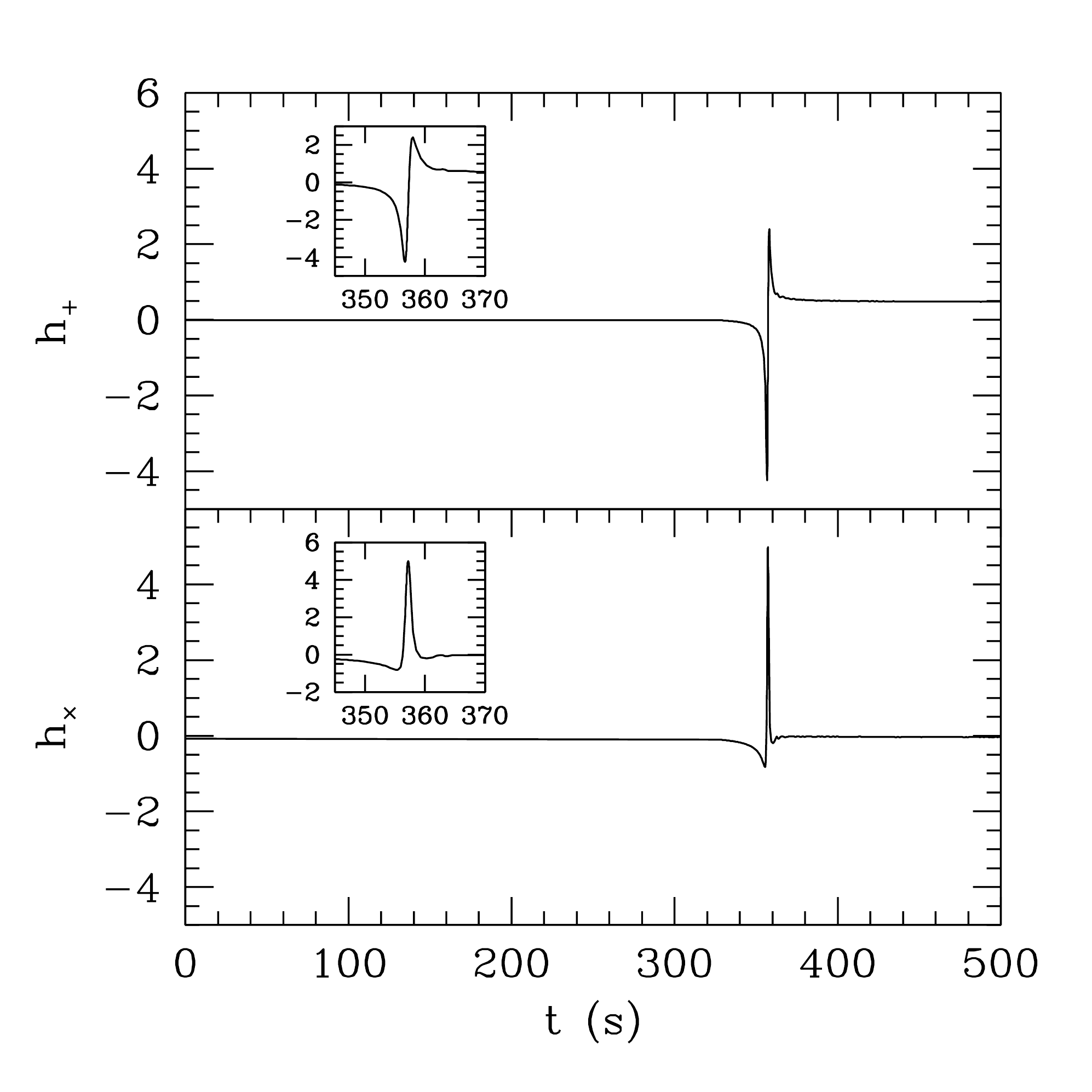}}
   \caption{Gravitational waveform for a  typical direct collision, in
     units of $10^{-21}$. The inset shows in greater detail the region
     of the burst of gravitational waves.}
\label{fig:DC:GWR}
\end{figure}

As explained  before, the last  outcome of the  dynamical interactions
studied here  consists in a direct  collision, in which a  single mass
transfer  episode,  of  very  short duration,  occurs.   The  physical
conditions  achieved  in all  these  interactions  are such  that  the
densities and temperatures necessary  to produce a powerful detonation
are met, leading  in nearly all the cases to  the disruption of either
the  lightest  white dwarf  or  of  both stars  ---  see  column 3  of
table~\ref{tab:DC:SNR}, where  we list whether  none (0), one  (1), or
both (2) colliding white dwarfs are disrupted --- and a sizable amount
of mass is  ejected as a consequence of the  interaction.  Although it
will not  be possible to  detect the gravitational signal  produced in
these interactions, owing  to their extremely short  duration, for the
sake of completeness in Fig.~\ref{fig:DC:GWR} a typical example of the
gravitational wave  pattern is displayed. The  signal is characterized
by a single, very narrow pulse.  Finally, the total energy radiated as
gravitational waves can be found  in table~\ref{tab:DC:SNR}. As can be
seen, in general, the larger the  total mass of the system, the larger
the  emission  of gravitational  waves.   As  it occurs  with  lateral
collisions ---  see Sect.~\ref{sec:GWR:LC} ---  the mass ratio  of the
colliding white dwarfs  also plays a role, and white  dwarf pairs with
smaller values of  $q$ release smaller amounts  of gravitational waves
for the same value of $M_1+M_2$.

\begin{table*}
\caption{Properties of the direct collisions.}
\label{tab:DC:SNR}
\begin{center}
\small
\begin{tabular}{ccccccc}
\hline 
\hline 
\noalign{\smallskip}
 Run &   $M_1+M_2$  & Disruption & $M_{\rm Ni}$  & $E_{\rm GW}$ & $E_{\nu}$  & $\frac{\Delta m}{\Delta t}$         \\
     & ($M_{\sun}$) &            & ($M_{\sun}$)  & \multicolumn{2}{c}{(erg)} & (mag/day)                           \\
\noalign{\smallskip}
\hline
\hline
\multicolumn{6}{l}{$v_{\rm ini} = 75 \;\text{km/s} \quad \quad \Delta y = 0.3 \;R_{\sun}$} \\
\hline
\noalign{\smallskip}
37 & 0.4+0.2 & 1 & 1.31$\times 10^{-9}$ & 9.51$\times 10^{37}$ & 3.21$\times 10^{43}$ & 0.019 \\
38 & 0.4+0.4 & 2 & 8.84$\times 10^{-4}$ & 1.49$\times 10^{41}$ & 7.01$\times 10^{45}$ & 0.016 \\
\hline
\multicolumn{6}{l}{$v_{\rm ini} = 75 \;\text{km/s} \quad \quad \Delta y = 0.4 \;R_{\sun}$} \\
\hline
\noalign{\smallskip}
1  & 0.8+0.6 & 0 & 8.65$\times 10^{-8}$ &  1.72$\times 10^{41}$ & 1.20$\times 10^{45}$ & ---  \\
2  & 0.8+0.8 & 0 & 4.47$\times 10^{-3}$ &  5.12$\times 10^{41}$ & 7.21$\times 10^{45}$ & ---  \\
3  & 1.0+0.8 & 2 & 7.25$\times 10^{-1}$ &  8.46$\times 10^{41}$ & 8.42$\times 10^{47}$ & 0.020 \\
4  & 1.2+0.8 & 1 & 6.60$\times 10^{-2}$ &  2.71$\times 10^{42}$ & 3.04$\times 10^{46}$ & 0.019 \\
40 & 0.4+0.4 & 2 & 1.64$\times 10^{-3}$ &  6.38$\times 10^{40}$ & 6.43$\times 10^{45}$ & 0.010 \\
41 & 0.8+0.4 & 1 & 8.04$\times 10^{-4}$ &  5.37$\times 10^{39}$ & 3.66$\times 10^{45}$ & 0.013 \\
42 & 1.2+0.4 & 1 & 1.18$\times 10^{-3}$ &  3.47$\times 10^{40}$ & 4.49$\times 10^{45}$ & 0.022 \\
\hline
\multicolumn{6}{l}{$v_{\rm ini} = 100 \;\text{km/s} \quad \quad \Delta y = 0.3 \;R_{\sun}$} \\
\hline
\noalign{\smallskip}
5  & 0.8+0.6 & 0 & 2.74$\times 10^{-8}$ & 1.70$\times 10^{41}$ & 1.03$\times 10^{45}$ & --- \\
6  & 0.8+0.8 & 0 & 3.67$\times 10^{-3}$ & 5.35$\times 10^{41}$ & 6.64$\times 10^{45}$ & --- \\
7  & 1.0+0.8 & 2 & 7.15$\times 10^{-1}$ & 9.08$\times 10^{41}$ & 8.36$\times 10^{47}$ & 0.017 \\
8  & 1.2+0.8 & 1 & 6.32$\times 10^{-2}$ & 2.62$\times 10^{42}$ & 3.18$\times 10^{46}$ & 0.013 \\
44 & 0.4+0.4 & 2 & 4.68$\times 10^{-3}$ & 9.33$\times 10^{40}$ & 6.94$\times 10^{45}$ & 0.020 \\
45 & 0.8+0.4 & 1 & 7.60$\times 10^{-4}$ & 5.73$\times 10^{39}$ & 3.66$\times 10^{45}$ & 0.015 \\
46 & 1.2+0.4 & 1 & 2.21$\times 10^{-3}$ & 3.39$\times 10^{40}$ & 4.95$\times 10^{45}$ & 0.015 \\
\hline
\multicolumn{6}{l}{$v_{\rm ini} = 100 \;\text{km/s} \quad \quad \Delta y = 0.4 \;R_{\sun}$} \\
\hline
\noalign{\smallskip}
49 & 0.8+0.4 & 1 & 5.00$\times 10^{-4}$ & 9.54$\times 10^{39}$ & 3.02$\times 10^{45}$ & 0.017 \\
50 & 1.2+0.4 & 1 & 9.36$\times 10^{-4}$ & 4.52$\times 10^{40}$ & 3.67$\times 10^{45}$ & 0.019 \\
\hline
\multicolumn{6}{l}{$v_{\rm ini} = 150 \;\text{km/s} \quad \quad \Delta y = 0.3 \;R_{\sun}$} \\
\hline
\noalign{\smallskip}
53 & 0.8+0.4 & 1 & 2.00$\times 10^{-4}$ & 1.51$\times 10^{40}$ & 2.47$\times 10^{45}$ & 0.015 \\
54 & 1.2+0.4 & 1 & 8.44$\times 10^{-4}$ & 5.11$\times 10^{40}$ & 3.39$\times 10^{45}$ & 0.018 \\
\noalign{\smallskip}
\hline
\hline
\end{tabular}
\end{center}
\end{table*}

\subsection{Light curves}
\label{sec:results:LCs}

\begin{figure}
   \resizebox{\hsize}{!}
   {\includegraphics[width=\columnwidth]{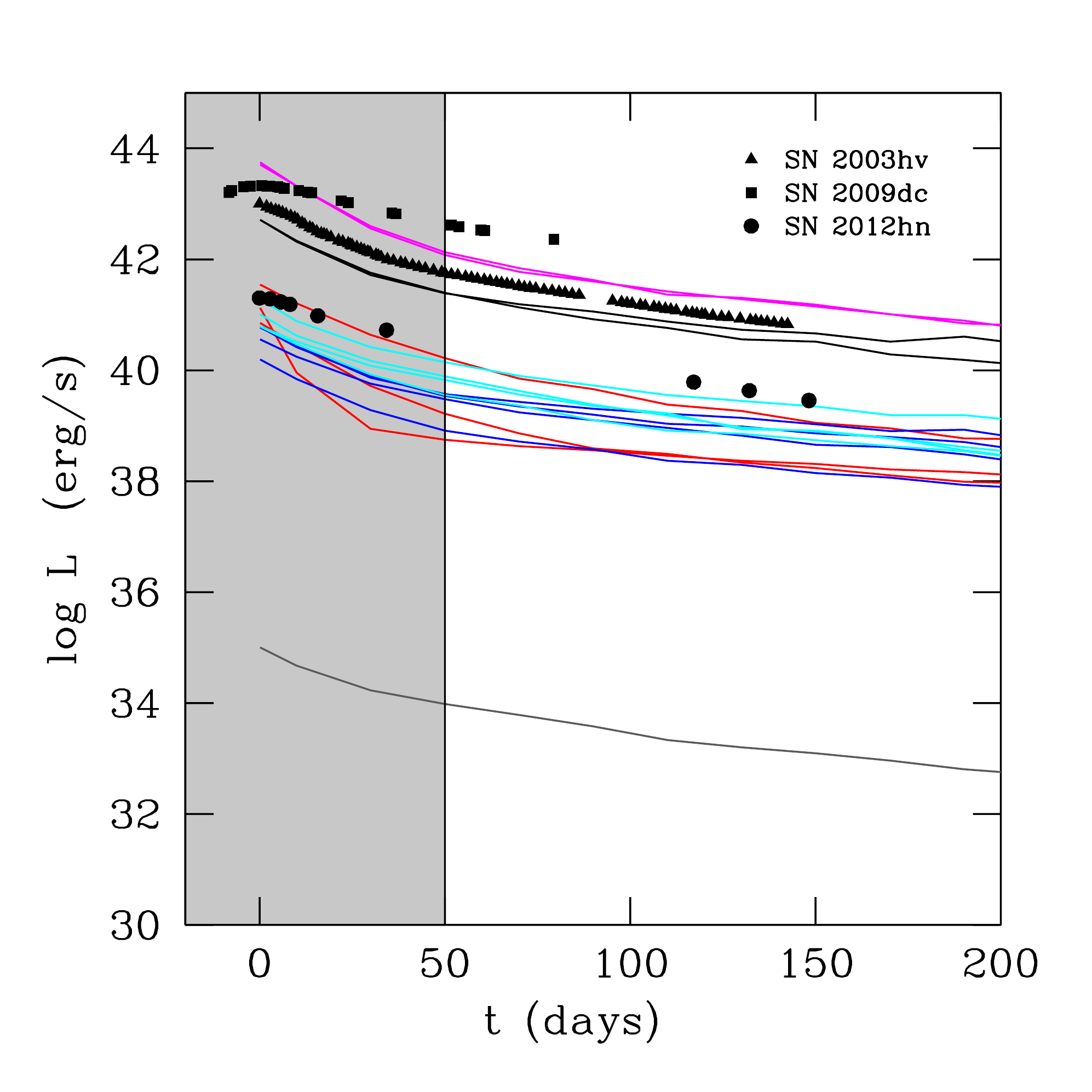}}
   \caption{Bolometric  late-time  light  curves  of  those  dynamical
     interactions in which at least  one of the colliding white dwarfs
     explodes.  The  colors indicate  the different white  dwarf pairs
     considered    here,    and    is    the    same    employed    in
     Fig.~\ref{fig:eo:elisa}:   red   curves   stand  for   a   $0.4\,
     M_{\sun}+0.4\,M_{\sun}$   system,  black   ones   for  a   $0.8\,
     M_{\sun}+1.2\,   M_{\sun}$,  magenta   curves  denote   a  $0.8\,
     M_{\sun}+1.0\,  M_{\sun}$ pair,  blue ones  correspond to  $0.4\,
     M_{\sun} +  0.8\, M_{\sun}$  systems, cyan lines  are used  for a
     $0.4\,  M_{\sun} +  1.2\, M_{\sun}$  pair, while  the grey  curve
     corresponds to the  $0.2\, M_{\sun} + 0.4\,  M_{\sun}$ case.  The
     light  curves  of three  observed  Type  Ia supernovae  are  also
     displayed for comparison.  Square symbols show the light curve of
     SN~2009dc,  triangles  correspond  to  SN~2003hv,  while  circles
     correspond to SN~2012hn.  The gray shaded area indicates that for
     these early times  our light curve is not  entirely reliable. The
     time origin is set at the time at which our simulations reach the
     maximum temperature (explosion time).  See the on-line edition of
     the journal for a color version of this figure.}
\label{fig:LCs}
\end{figure}

As some of the white dwarf dynamical interactions analyzed here result
in  powerful explosions,  the light  curves  powered by  the decay  of
radioactive $^{56}$Ni  synthesized in the  most violent phases  of the
interaction --- when the material  is shocked and the temperatures and
densities are such  that a detonation is able to  develop --- might be
eventually  detectable.  As  extensively discussed  before, explosions
are  more likely  to  occur  in direct  collisions,  although in  some
lateral  collisions  some nickel  is  also  synthesized. However,  the
masses  of synthesized  nickel in  lateral collisions  are in  general
small, and consequently  we expect that most of these  events would be
undetectable --- see  tables~\ref{tab:LC:SNR} and \ref{tab:DC:SNR}. In
particular the amount  of $^{56}$Ni produced in  lateral collisions is
almost negligible in  most cases --- ranging from  about $\sim 5\times
10^{-14}\, M_{\sun}$  to about  $1\times 10^{-5}\, M_{\sun}$.   On the
other  hand, the  resulting  masses of  nickel  in direct  collisions,
although show a broad range of variation, are consistently larger than
in lateral collisions.  According to  this, the late-time light curves
are   characterized   by  large   variations.    This   is  shown   in
Fig.~\ref{fig:LCs},  where the  late-time  light  curves ---  computed
according to the procedure outlined in Sect.~\ref{sec:EM} --- of those
direct collisions in which at least  one of the colliding white dwarfs
is disrupted are displayed.

Fig.~\ref{fig:LCs} shows that the late-time light curves are sensitive
not only to the total mass of  the pair of colliding white dwarfs, but
also to the choice of initial conditions. Specifically, the bolometric
late-time light  curves can  differ considerably for  a fixed  pair of
masses,  depending  on  the   initial  conditions  of  the  considered
interaction.  The most extreme case is that of a $0.4\, M_{\sun}+0.4\,
M_{\sun}$ (red lines in  Fig.~\ref{fig:LCs}).  In particular, for this
specific case the late-time bolometric  luminosities differ by about 1
order of  magnitude when  runs number  38, 40  and 44  are considered.
This  stems  from the  very  large  difference  in the  nickel  masses
synthesized  in the  respective  interactions,  which are  $8.84\times
10^{-4}\, M_{\sun}$, $1.64\times  10^{-3}\, M_{\sun}$, and $4.68\times
10^{-3}\,  M_{\sun}$, respectively.  The different  velocities of  the
ejecta also  play a role.   The most powerful outburst  corresponds to
the  case in  which a  $0.8\, M_{\sun}+1.0\,  M_{\sun}$ pair  of white
dwarfs   experience  a   direct   collision  ---   magenta  lines   in
Fig.~\ref{fig:LCs}. This agrees well with  the results obtained so far
for white  dwarf mergers, in which  case the most powerful  events are
found when  both white dwarfs  have similar masses and  are relatively
massive \citep{Pakmor2012}.   Also of interest  is to realize  that we
obtain a mild explosion in the case in which a $0.2\, M_{\sun} + 0.4\,
M_{\sun}$ white  dwarf system  in which both  stars have  helium cores
collide (run  number 37).  In  this specific  case the nickel  mass is
small ($1.39\times 10^{-9}\, M_{\sun}$), and only one of the stars ---
the lightest  one ---  is disrupted.  Most importantly,  its late-time
light curve  falls well below the  bulk of light curves  computed here
(the peak luminosity is nearly 9 orders of magnitude smaller).

In Fig.~\ref{fig:LCs} we  also display the bolometric  light curves of
three thermonuclear  supernovae, for which late-time  observations are
available,  to  allow  a  meaningful  comparison  with  the  synthetic
late-time light curves computed here.  Squares correspond to the light
curve  of  SN~2009dc, which  was  an  overluminous peculiar  supernova
\cite{2009dc},  triangles  correspond  to  the  light  curve  of  the
otherwise  normal  thermonuclear supernova  SN~2003hv  \citep{2003hv},
while  circles   correspond  to   the  underluminous   SNIa  SN~2012hn
\citep{PESSTO}.  As can be seen, only a few of our simulated late-time
light  curves  --- corresponding  to  the  brightest events  ---  have
characteristics similar to those  of thermonuclear supernovae, whereas
most of the dynamical interactions  studied in this paper would appear
as underluminous  transients. In  particular, only  the runs  in which
rather massive white dwarfs interact  have light curves which could be
assimilated to those of SNIa. In  particular, this is the case of runs
number 4  and 8, in  which a pair of  white dwarfs with  masses $0.8\,
M_{\sun}$ and  $1.2\, M_{\sun}$  interact --- in  which case  only the
less massive carbon-oxygen white dwarf is disrupted, while the massive
ONe one remains bound --- for runs 3 and 7 --- which involve a pair of
massive  carbon-oxygen white  dwarfs  of masses  $0.8\, M_{\sun}$  and
$1.0\, M_{\sun}$, and  both components are distroyed  as a consequence
of the  interaction --- and,  possibly, runs  number 38 and  42, which
involve  two  white  dwarfs  with masses  $0.4\,M_{\sun}$  and  $0.4\,
M_{\sun}$, and  $1.2\, M_{\sun}$  and $0.4\,  M_{\sun}$, respectively.
Note that in these last two cases  a white dwarf with a helium core is
involved. Hence,  the total mass  burned is small,  as is the  mass of
synthesized  $^{56}$Ni ---  see  table~\ref{tab:DC:SNR}.  In  summary,
only those light curves resulting from  events in which the total mass
of  the system  is large  (larger than  $\sim 1.8\,  M_{\sun}$) and  a
carbon-oxygen  white dwarf  is  involved in  the interaction  resemble
those   of  thermonuclear   supernovae,  whereas   the  rest   of  the
interactions are transient bright events of different nature.

In  table~\ref{tab:DC:SNR}  we  also  show the  decline  rate  of  the
bolometric light curve  at late times.  As well known,  after about 50
days the light curves of SNIa  steadily decline in an exponential way.
It is observationally found that the decline rates are essentially the
same  for  all  thermonuclear  supernovae  between  50  and  120  days
\citep{Wells,Hamuy,Lira} --- that is, for  the time interval for which
the  results  of  our  calculations  are reliable  ---  and  hence,  a
comparison  of our  calculations  and the  observed  decline rates  is
worthwhile.   The typical  values of  the late-time  decline rate  are
0.014~mag/day  in the  $B$ band,  0.028~mag/day in  the $V$  band, and
0.042~mag/day in  the $I$  band, although there  are a  few exceptions
(SN~1986G   and    SN~1991bg),   which   decline   faster    ---   see
\cite{Leibundgut},  and  references   therein.   Given  our  numerical
approach, to  compare our  simulated late-time  light curves  with the
observed decline  rates of SNeIa  we only used  data from day  60.  As
shown  in  table~\ref{tab:DC:SNR}  the   computed  decline  rates  are
compatible with those observationally found in Type Ia supernovae.

\subsection{Thermal neutrinos}
\label{sec:results:nu}

\begin{figure}
   \resizebox{\hsize}{!}
   {\includegraphics[width=\columnwidth]{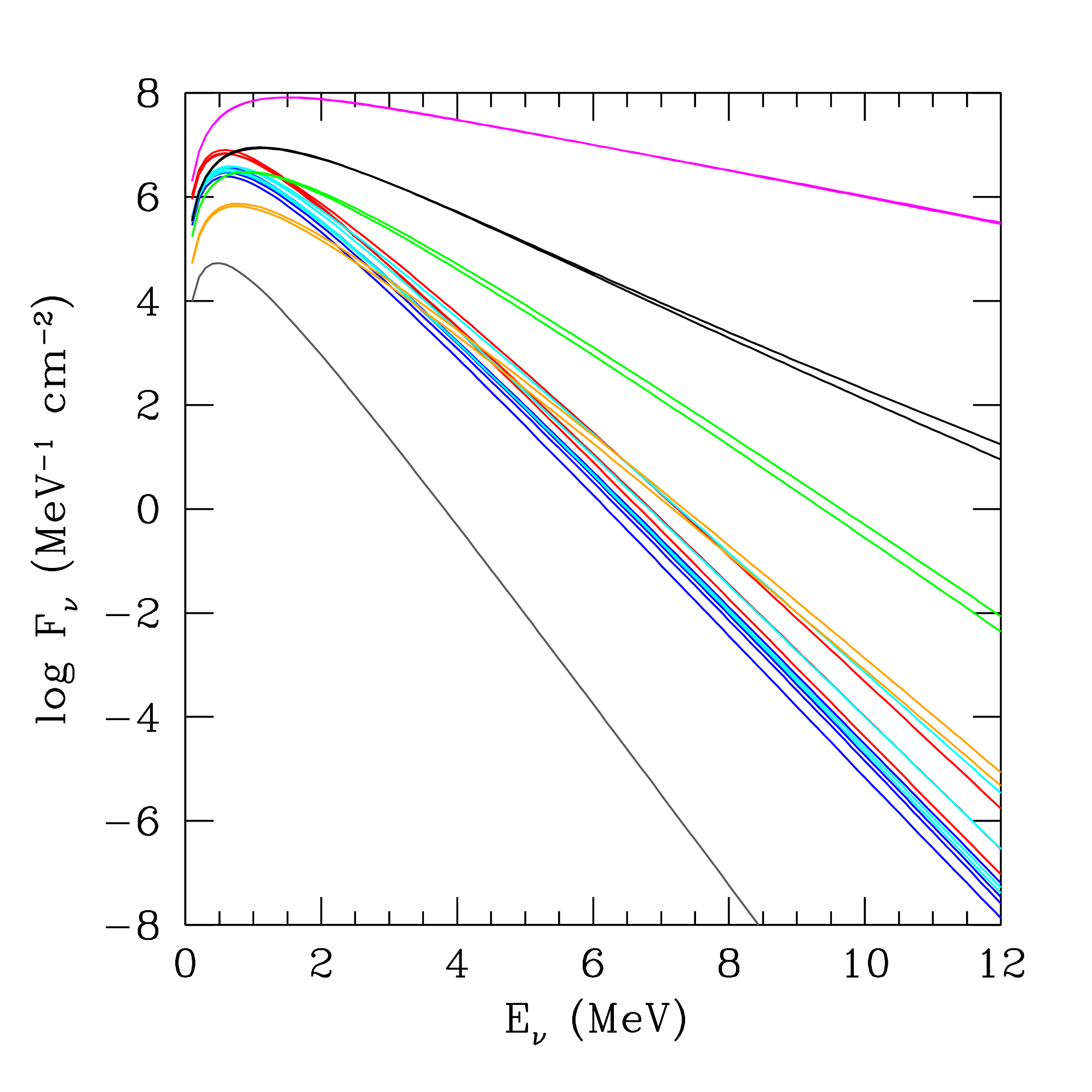}}
   \caption{Spectral  energy  distribution   of  neutrinos  for  those
     interactions resulting  in a direct collision.   The color coding
     is the same employed in Fig.~\ref{fig:eo:elisa}.  We assumed that
     source of neutrinos  is located at a distance of  1~kpc.  See the
     on-line edition for a color version of this figure.}
\label{fig:SEDnu}
\end{figure}

As mentioned, due  to the extreme physical conditions  reached in some
of the interactions  studied in this work it is  expected that copious
amounts of thermal neutrinos should be produced in those regions where
the material of  the colliding white dwarfs is  compressed and heated.
Tables~\ref{tab:LC:SNR} and \ref{tab:DC:SNR} show  that this is indeed
the case. In  particular, in these tables we list  the energy radiated
in the form  of thermal neutrinos in lateral collisions  and in direct
ones, respectively.  As can be seen,  the radiated energies are in all
cases   rather    large,   although    a   large   spread    is   also
found. Specifically, for  the case of lateral  collisions the neutrino
energies range  from $\sim  1.7\times 10^{35}$~erg to  $\sim 2.4\times
10^{45}$~erg, roughly 10  orders of magnitude, whereas in  the case of
direct collisions the range of variation is somewhat smaller, since it
spans about 5 orders of magnitude, being run 3 the simulation in which
more  neutrinos  are  produced  ($\sim  8.4\times  10^{47}$~erg).   In
general, we  find that for  the case of lateral  collisions increasing
the total  mass of  the system results  in stronger  interactions, and
consequently  larger peak  temperatures  and  densities are  achieved.
This directly  translates in a  larger release of both  nuclear energy
and of thermal neutrinos --- see table~\ref{tab:LC:SNR}.  Equally, for
a fixed total mass of the system, larger mass ratios lead to a smaller
nuclear energy release, and in  a reduced neutrino emission.  Besides,
systems with larger initial distances at closest approach between both
colliding white dwarfs result in longer  mergers --- as can be seen in
column 3  of table~\ref{tab:LC:SNR},  which lists  the number  of mass
transfer  episodes  occurring  in  each  lateral  collision  ---  and,
therefore,  in  more  gentle  interactions, thus  leading  to  smaller
neutrino luminosities.  This is in contrast with what we found for the
emission of  gravitational waves, for  which the reverse is  true. For
the case  of direct collisions we  also find that, in  general, as the
total  mass of  the system  is increased,  neutrino luminosities  also
increase,   as   it   happens    in   lateral   collisions   ---   see
table~\ref{tab:DC:SNR}.  However,  there are exceptions to  this rule.
Specifically, a good  example of this is the case  of the interactions
in  which two  white  dwarfs  of masses  $0.8\,  M_{\sun}$ and  $1.0\,
M_{\sun}$  collide.   In these  interactions  both  stars are  totally
disrupted, and  it turns  out that,  consequently, the  nuclear energy
released is larger,  and that more neutrinos are produced  than in the
case in which a $0.8\, M_{\sun} + 1.2\, M_{\sun}$ is considered.  This
is  a consequence  of the  fact  that in  this case  the more  massive
oxygen-neon  white   dwarf  of   $1.2\,  M_{\sun}$  is   more  tightly
gravitationally bound, and is not disrupted during the interaction.

\begin{figure}
   \resizebox{\hsize}{!}
   {\includegraphics[width=\columnwidth]{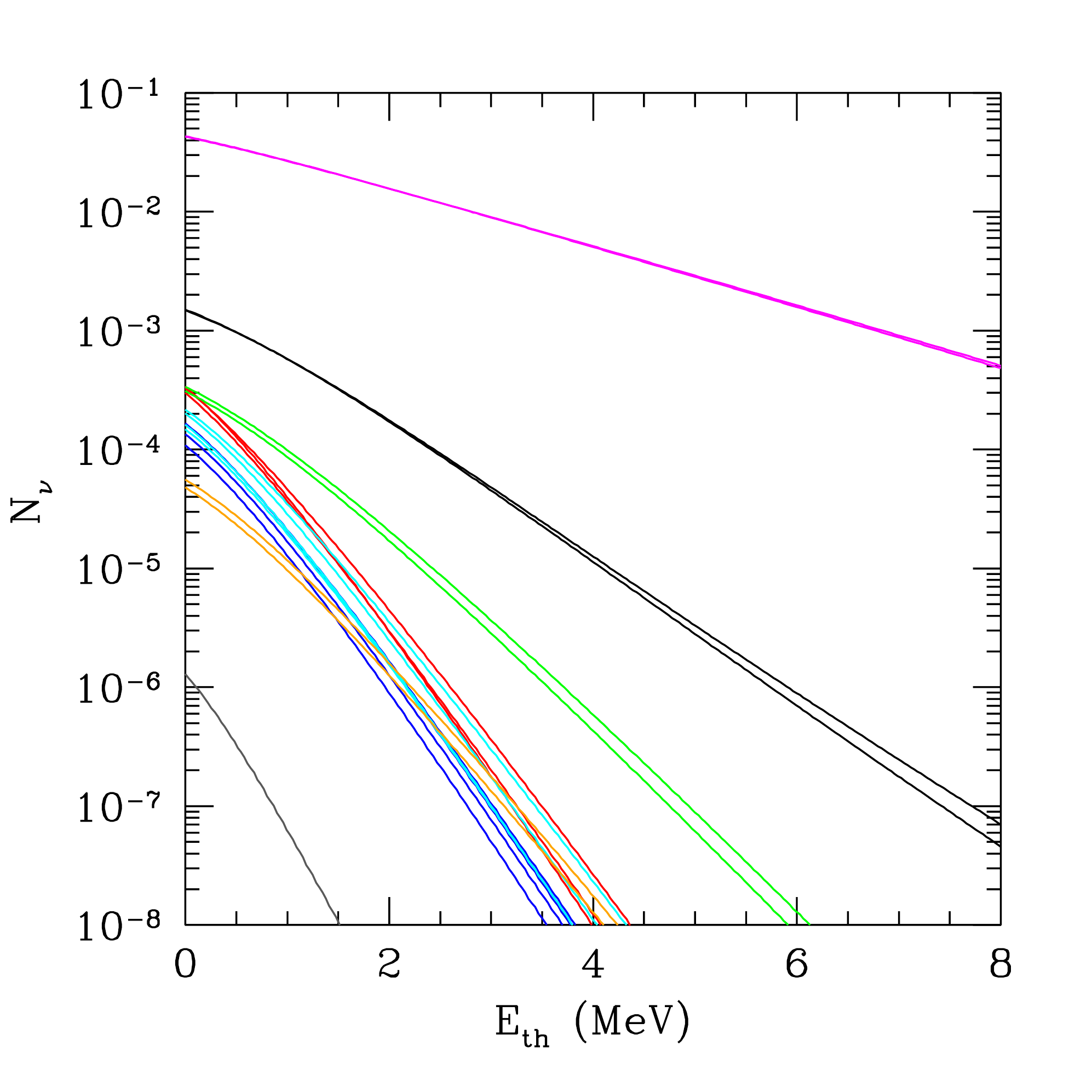}}
   \caption{Expected number of neutrino events in the Super-Kamiokande
     detector,  when the  source is  located at  a distance  of 1~kpc.
     Again,   the    color   coding   is   the    same   employed   in
     Fig.~\ref{fig:LCs}. See the on-line edition  of the journal for a
     color version of this figure.}
\label{fig:S-K}
\end{figure}

Figure \ref{fig:SEDnu}  displays the  spectral energy  distribution of
the neutrinos emitted  when the temperature reaches  its maximum value
during the  dynamical interaction, using  the same color coding  as it
was done in previous figures. As expected, the general features of the
neutrino spectral  flux follow the  same pattern previously  found for
the  peak   bolometric  luminosities,  and   show  a  wide   range  of
variation. Quite naturally, the simulations in which more $^{56}$Ni is
synthesized also produce more neutrinos. Nevertheless, this is not the
most   relevant   information  that   can   be   obtained  from   this
figure. Instead, it is important to note that in all cases the peak of
emission corresponds to  neutrinos with energies between  1 and 2~MeV.
Thus,  the best-suited  detector  for searching  the neutrino  signals
produced in  these dynamical  events is the  Super-Kamiokande detector
\citep{Nakahata}, for  which the  peak sensitivity is  3~MeV, although
other detectors, like the IceCube Observatory \citep{Ahrens2002} could
also eventually  detect the  neutrinos produced in  these interactions
\citep{Abbasi2011}.  Hence, we compute the number of expected neutrino
events  for   this  detector,  using  the   prescription  outlined  in
Sect.~\ref{sec:nu}.   The  results  are shown  in  Fig.~\ref{fig:S-K}.
Clearly,  even  for  nearby  sources (located  at  1~kpc)  only  those
dynamical interactions  in which two very  massive carbon-oxygen white
dwarfs are involved will have (small) chances to be detected, since in
the best of the cases only $\sim 0.04$ thermal neutrinos will interact
with the tank.

\subsection{Fallback luminosities}
\label{sec:results:Xray}

\begin{figure}
   \resizebox{\hsize}{!}
   {\includegraphics[width=\columnwidth]{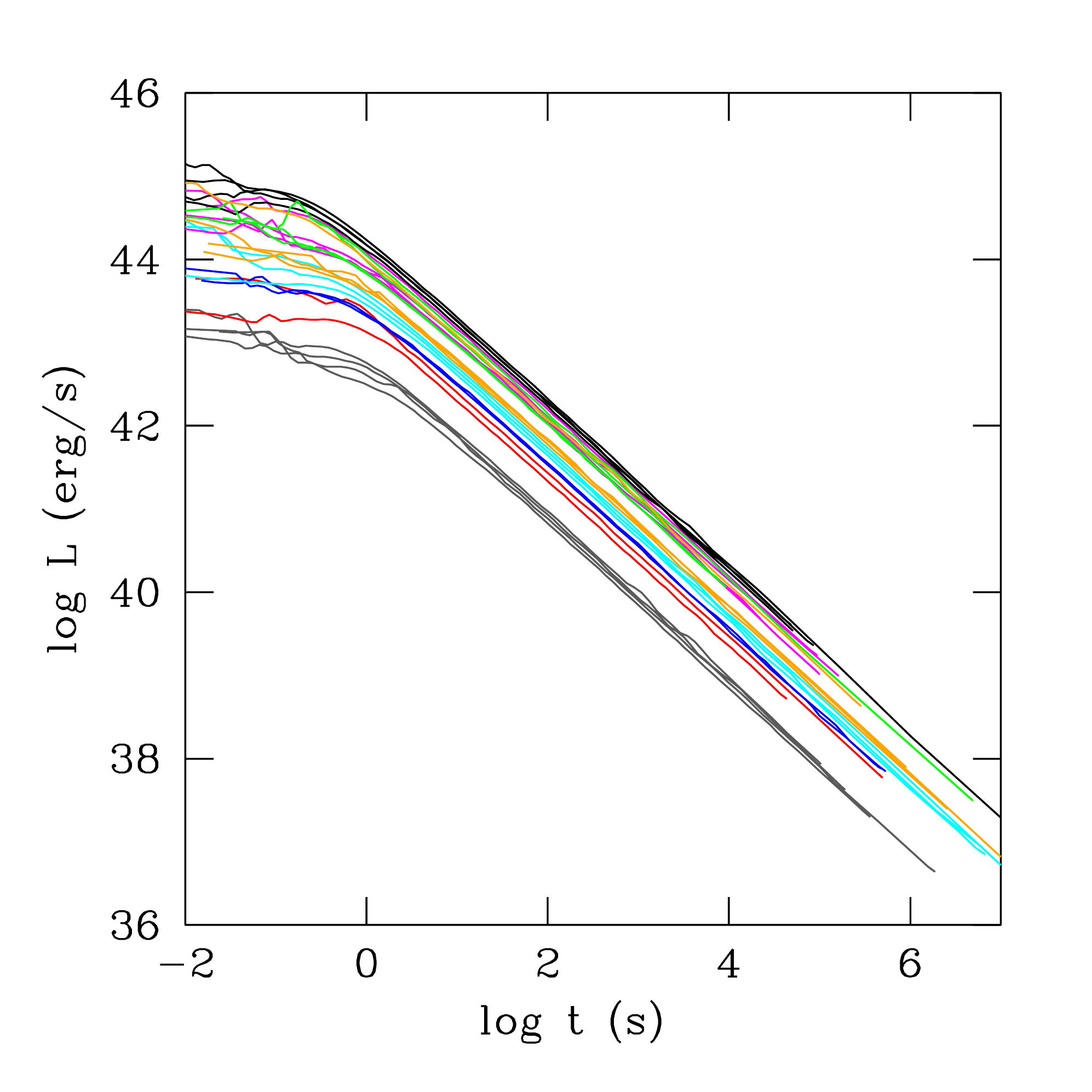}}
   \caption{Fallback   accretion   luminosity    for   the   dynamical
     interactions for  which the outcome  is a lateral  collision, and
     hence result in the formation  of a central remnant surrounded by
     a disk.   The fallback accretion luminosities  have been averaged
     over  all the  SPH particles  with eccentric  orbits for  a given
     time.  The color  coding is the same  of Fig.~\ref{fig:LCs}.  See
     the on-line  version of the journal  for a color version  of this
     figure.}
\label{fig:fall}
\end{figure}

Another potential observational signature  of the mergers studied here
is the emission  of high-energy photons from the  fallback material in
the  aftermath of  those dynamical  interactions in  which one  of the
stars is disrupted, and the  final configuration consists of a central
remnant  surrounded by  a debris  region.   As described  in depth  in
\cite{Aznar2013},  in some  dynamical interactions  a fraction  of the
disrupted star  goes to form a  Keplerian disk. In this  debris region
the vast majority  of material has circularized  orbits. However, some
material in this region has  highly eccentric orbits. After some time,
this material will most likely interact with the recently formed disk.
\cite{Rosswog2007} demonstrated  that the  relevant timescale  in this
case  is  not  given  by  viscous dissipation  but,  instead,  by  the
distribution of eccentrities. As discussed in Sect.~\ref{sec:fall}, we
adopted  the  formulation  of \cite{Rosswog2007}  and  calculated  the
accretion luminosity  resulting from  the interaction of  the material
with high eccentricities  with the newly formed disk  by assuming that
the kinetic energy of these  particles is dissipated within the radius
of the debris disk.

Figure~\ref{fig:fall} displays  the results of this  procedure for all
the  lateral  collisions  in  which  as  a  result  of  the  dynamical
interaction the less massive star is disrupted and a Keplerian disk is
formed.  We note that only a  fraction of the fallback luminosity will
be released  in the  form of high-energy  photons.  Thus,  the results
shown in this figure can be regarded  as an upper limit for the actual
luminosity of high-energy  photons.  It is important to  note that the
time dependence of  the fallback luminosities is very  similar for all
the simulations, $\propto t^{-5/3}$. This  time dependence is the same
found for double white dwarf  or double neutron star mergers.  Another
interesting point is  that the fallback luminosities  are rather high,
of the  order of $10^{45}$~erg~s$^{-1}$,  and thus these  events could
eventually be detected up to  relatively large distances.  Finally, as
it should be expected, the  most violent interactions result in larger
fallback  luminosities.  In  particular, the  events with  the largest
fallback luminosities are those in  which two white dwarfs with masses
$0.8\, M_{\sun}$ and $1.2\, M_{\sun}$ interact (runs number 12, 16, 20
and 24,  black lines),  while the  event in  which less  X-ray photons
would be radiated  is that in which  a pair of helium  white dwarfs of
masses $0.2\, M_{\sun}$ and $0.4\,  M_{\sun}$ collide (runs number 39,
43, 47,  and 51, grey  lines).  Note, nevertheless, that  the fallback
luminosity in this  case is only a factor of  $10^2$ smaller than that
computed for  the most violent  event.  This  is in contrast  with the
large difference  (almost 9 orders  of magnitude) found for  the light
curves of  direct collisions.   This, again, is  a consequence  of the
very  small amount  of $^{56}$Ni  synthesized during  the interaction,
whereas the  masses of the  debris region do not  differ substantially
(being  of the  same order  of magnitude).   A more  detailed analysis
shows that  the key issue  to explain this  behavior is that  the more
massive mergers produce more material in the debris region with larger
kinetic energies, thus resulting in enhanced fallback luminosities.

\section{Summary and conclusions}
\label{sec:conclusions}

We  have  computed  the  observational  signatures  of  the  dynamical
interactions of two white dwarfs  in a dense stellar environment. This
includes the emission  of gravitational waves for  those systems which
as a  result of  the interaction  end up  forming an  eccentric binary
system, or those pairs which  experience a lateral collision, in which
several mass  transfer episodes occur,  although we also  computed the
emission of  gravitational waves  for those events  in which  a direct
collision, in which only one violent mass transfer episode occurs. For
those cases in  which as a consequence of the  dynamical mass transfer
process  an  explosion  occurs,  and  either one  or  both  stars  are
disrupted,  we also  computed the  corresponding late-time  bolometric
light curves, and the associated  emission of thermal neutrinos, while
for all  those simulations in  which the  less massive white  dwarf is
disrupted, and part of its mass goes  to form a debris region, we also
computed the  fallback luminosities radiated  in the aftermath  of the
interaction.   This has  been  done employing  the  most complete  and
comprehensive  set of  simulations of  this  kind ---  namely that  of
\cite{Aznar2013} ---  which covers  a wide  range of  masses, chemical
compositions of the cores of  the white dwarfs, and initial conditions
of the intervening  stars. For all these signals we  have assessed the
feasibility of detecting them.

We  have  shown that  in  the  case  of  interactions leading  to  the
formation of  an eccentric binary  the most noticeable feature  of the
emitted gravitational  wave pattern is  a discrete spectrum,  and that
these  signals  are  likely  to   be  detected  by  future  spaceborne
detectors, like  eLISA, up to  relatively long distances  (larger than
10~kpc).  This, however,  is not  the case  of the  gravitational wave
radiation resulting  from lateral  collisions, although  more advanced
experiments, like ALIA,  would be able to detect  them. Finally, since
for the  case of direct  collisions the  emitted signal consists  of a
single, well-defined peak, of very  short duration, there are no hopes
to detect them.

The  late-time bolometric  light curves  of those  events in  which an
explosion  is able  to  develop, and  at  least one  of  the stars  is
disrupted, show  a broad range  of variation,  of almost 10  orders of
magnitude in luminosity. This is  the logical consequence of the large
variety of masses of $^{56}$Ni synthesized during the explosion. This,
in turn,  can be explained by  the very different masses  of the white
dwarfs involved in the interaction.  Even more, we have found that for
a given pair of white dwarfs  with fixed masses the initial conditions
of the interaction --- namely,  the initial separations and velocities
(or, alternatively, the energies and  angular momenta) --- also play a
key role  in shaping the  late-time bolometric light curves,  and that
the corresponding peak  luminosities can differ by almost  2 orders of
magnitude.  More interestingly, we have  found that only the brightest
events have light curves resembling those of thermonuclear supernovae,
and that most of our  simulations result in very underluminous events,
which would most likely be classified as bright transient events. Only
those  events in  which two  rather  massive white  dwarfs (of  masses
larger than  $\sim 0.8\,  M_{\sun}$) collide  would have  light curves
which could  be assimilated to those  of Type Ia supernovae.   Even in
this  case, the  resulting late-time  bolometric light  curves show  a
considerable  range of  variability depending  on the  adopted initial
conditions, and light curves resembling those of underluminous, normal
and peculiar bright SNIa are possible.

The corresponding thermal neutrino luminosities also show a noticeable
dispersion,  which is  a  natural consequence  of  the very  different
maximum temperatures  reached during  the most  violent phases  of the
interaction.   Nonetheless, the  chances  of  detecting the  neutrinos
emitted in these  events are very low for the  current detectors, even
if the dynamical  interaction occurs relatively close,  at 1~kpc. Even
in  this case,  the Super-Kamiokande  detector  would not  be able  to
detect the neutrino  signal, since the number counts  are small, $\sim
10^{-2}$ in the best of the cases.

Finally, we have also computed  the emission of high-energy photons in
the aftermath of those interactions in which at least one of the white
dwarfs is disrupted, and a debris  region is formed.  As it occurs for
the case of the mergers of two  white dwarfs, or of two neutron stars,
the accretion luminosity  follows a characteristic power  law of index
$-5/3$, but the fallback luminosities are considerably smaller than in
the  case of  neutron star  mergers.  Nevertheless,  the typical  peak
luminosities  are of  the order  of $10^{44}$~erg~s$^{-1}$,  and hence
should be easily detectable up to  very long distances. Again, we also
find that  depending on the masses  and the initial conditions  of the
interaction the  spread in  the peak  fallback luminosities  is large,
although  considerably smaller  than that  obtained for  the late-time
bolometric light curves.

All  in all,  our  calculations provide  a qualitative  multimessenger
picture of the dynamical interactions of two white dwarfs.  A combined
strategy  in   which  data   obtained  from  the   planned  spaceborne
gravitational  wave  detectors, as  well  as  optical and  high-energy
observations would result in valuable  insight on the conditions under
which this type of events take place.

\section*{Acknowledgments}
This work  was partially supported  by MCINN grant  AYA2011--23102, by
the  AGAUR and  by  the  European Union  FEDER  funds. We  acknowledge
B. Katz and  D. Kurshnir for making available to  us their Monte Carlo
code to compute the light curves presented in this work. We also thank
L. Rezzolla  for helpful  discussions, and  our anonymous  referee for
valuable comments and suggestions.

\bibliographystyle{mn2e}
\bibliography{GWRc}

\end{document}